\documentclass[apj]{emulateapj}

\usepackage{graphicx}
\usepackage{subfigure}
\usepackage{color} %% red, green, blue (in GMKY, cyan, magenta, yellow)
\usepackage{longtable,lscape,rotating}
\usepackage{enumerate}
\newcommand{\uu}{$U$}
\newcommand{\bb}{$B$}
\newcommand{\vv}{$V$}

\newcommand{\iw}{$I$}
\newcommand{\ha}{H$\alpha$~}

\newcommand{\etal}{et~al.}

\newcommand{\eqref}[1]{equation~(\ref{#1})}

\newcommand{\hst}{$HST$}
\newcommand{\ubvi}{$UBVI$}
\newcommand{\vi}{$(V\!-\!I)$}
\newcommand{\uub}{$(U\!-\!B)$}

\begin{document}

\title{The Resolved Stellar Population in 50 Regions of M83 from HST/WFC3 Early Release Science Observations}

\shorttitle{RESOLVED STARS IN M\,83}

\author{Hwihyun Kim\altaffilmark{1}, 
Bradley C.~Whitmore\altaffilmark{2},
Rupali Chandar\altaffilmark{3},
Abhijit Saha\altaffilmark{4},
Catherine C.~Kaleida\altaffilmark{5},
Max Mutchler\altaffilmark{2},
Seth H.~Cohen\altaffilmark{1},
Daniela Calzetti\altaffilmark{6}, 
Robert W.~O'Connell\altaffilmark{7},
Rogier A.\ Windhorst\altaffilmark{1},
Bruce Balick\altaffilmark{8},
Howard E.~Bond\altaffilmark{2},
C.~Marcella Carollo\altaffilmark{9},
Michael J.~Disney\altaffilmark{10},
Michael A.~Dopita\altaffilmark{11,12},
Jay A.~Frogel\altaffilmark{13,14},
Donald N.~B.~Hall\altaffilmark{12},
Jon A.~Holtzman\altaffilmark{15},
Randy A.~Kimble\altaffilmark{16},
Patrick J.~McCarthy\altaffilmark{17},
Francesco Paresce\altaffilmark{18},
Joe I.~Silk\altaffilmark{19},
John T.~Trauger\altaffilmark{20},
Alistair R.~Walker\altaffilmark{5},
Erick T.~Young \altaffilmark{21}
}

\altaffiltext{1}{School of Earth and Space Exploration, Arizona
  State University, Tempe, AZ 85287-1404, USA} 
\altaffiltext{2}{Space Telescope Science Institute, Baltimore, MD 21218}
\altaffiltext{3}{Department of Physics \& Astronomy, University of
  Toledo, Toledo, OH 43606}
\altaffiltext{4}{National Optical Astronomy Observatories, Tucson, AZ
85726-6732} 
\altaffiltext{5}{Cerro Tololo Inter-American Observatory,
La Serena, Chile}
\altaffiltext{6}{Department of Astronomy, University of Massachusetts,
Amherst, MA 01003}
\altaffiltext{7}{Department of Astronomy, University of Virginia,
Charlottesville, VA 22904-4325}
\altaffiltext{8}{Department of Astronomy, University of Washington,
Seattle, WA 98195-1580}
\altaffiltext{9}{Department of Physics, ETH-Zurich, Zurich, 8093
Switzerland}
\altaffiltext{10}{School of Physics and Astronomy, Cardiff University,
 Cardiff CF24 3AA, United Kingdom}
\altaffiltext{11}{Mount Stromlo and Siding Spring Observatories,
Research School of Astronomy \& Astrophysics, 
Australian National University, Cotter Road,
Weston Creek, ACT 2611, Australia}
\altaffiltext{12}{Institute for Astronomy, University of Hawaii,
 2680 Woodlawn Drive, Honolulu, HI 96822}
\altaffiltext{13}{Galaxies Unlimited, 1 Tremblant Court, Lutherville, MD 21093}
\altaffiltext{14}{Astronomy Department, King Abdulaziz University, P.O. Box 80203, Jedda, Saudi Arabia}
\altaffiltext{15}{Department of Astronomy, New Mexico State
University, Las Cruces, NM 88003}
\altaffiltext{16}{NASA--Goddard Space Flight Center, Greenbelt, MD 20771}
\altaffiltext{17}{Observatories of the Carnegie Institution of
Washington, Pasadena, CA 91101-1292}
\altaffiltext{18}{European Southern Observatory, Garching bei M\"unchen,
85748 Germany }
\altaffiltext{19}{Department of Physics, University of Oxford, Oxford
OX1 3PU, United Kingdom}
\altaffiltext{20}{NASA--Jet Propulsion Laboratory, Pasadena, CA 91109}
\altaffiltext{21}{NASA--Ames Research Center, Moffett Field, CA 94035}

\email{hwihyun.kim@asu.edu}
\shortauthors{Kim et al.}

%--------------------------------------------------
%%%%%%%%%% Abstract %%%%%%%%%%
\begin{abstract}

We present a multi-wavelength photometric study of $\sim$15,000
resolved stars in the nearby spiral galaxy M83 (NGC\,5236,
$D$\,$=$\,$4.61$\,Mpc) based on $Hubble$ $Space$ $Telescope$ Wide
Field Camera 3 observations using four filters:\,F336W, F438W, F555W,
and F814W.  We select 50~regions (an average size of
260\,pc~by~280\,pc) in the spiral arm and inter-arm areas of M83, and
determine the age distribution of the luminous stellar populations in
each region.  This is accomplished by correcting for extinction
towards each individual star by comparing its colors with predictions
from stellar isochrones.  We compare the resulting luminosity weighted
mean ages of the luminous stars in the 50~regions with those
determined from several independent methods, including the number
ratio of red-to-blue supergiants, morphological appearance of the
regions, surface brightness fluctuations, and the ages of clusters in
the regions. We find reasonably good agreement between these methods.
We also find that young stars are much more likely to be found in
concentrated aggregates along spiral arms, while older stars are more
dispersed.  These results are consistent with the scenario that star
formation is associated with the spiral arms, and stars form primarily
in star clusters and then disperse on short timescales to form the
field population.  The locations of Wolf-Rayet stars are found to
correlate with the positions of many of the youngest regions,
providing additional support for our ability to accurately estimate
ages.  We address the effects of spatial resolution on the measured
colors, magnitudes, and age estimates. While individual stars can
occasionally show measurable differences in the colors and magnitudes,
the age estimates for entire regions are only slightly affected.

\end{abstract}

%%%%%%%%%% Subject headings %%%%%%%%%%
\keywords{galaxies: individual (M83, NGC\,5236) --- galaxies: stellar content} 
% --- galaxies: structure}

%--------------------------------------------------
%%%%%%%%%% S1 Introduction %%%%%%%%%%                            
\section{Introduction}

Understanding the properties of stars and the history of star
formation in galaxies remains one of the most fundamental subjects in
astrophysics. The {\it Hubble Space Telescope} (\hst) provides an
important tool for this endeavor, since it enables the detailed study
of stars and star clusters, not only in the Milky Way and its nearest
neighbors, but in galaxies well beyond the Local Group. The Wide Field
Camera 3 (WFC3) provides a particularly valuable new panchromatic
imaging capability, with spectral coverage from the near-UV to the
near-IR.  This capability is especially useful for studying the
stellar populations in nearby galaxies where individual stars are
resolved.  

A good example of a multi-wavelength survey of nearby galaxies which
employs resolved stars and star clusters is the ACS Nearby Galaxy
Treasury (ANGST) program \citep{dalcanton09}. This study includes
$\sim$65 galaxies out to $\sim$3.5~Mpc, and provides uniform
multi-color ($BVI$) catalogs of tens of thousands of individual stars
in each galaxy. Although the ANGST program provides an excellent survey
for a wide range of studies, it does not provide observations in the
\uu-band since the observations were obtained before WFC3 was
installed on \hst.  The \uu-band is particularly useful for age-dating
populations of young stars, which is the focus of the current paper.
PHAT (Panchromatic Hubble Andromeda Treasury) is a related study that
does take advantage of the new \uu-band capability on WFC3. While
nowhere as extensive as the Multiple Cycle Treasury Program PHAT
survey, the current study of M83 is complementary in the sense that it
provides similar observations of the resolved stellar component of a
nearby spiral galaxy, and uses quite different analysis techniques, as
will be discussed in Sections 2 and 3.  Future observations of M83
(ID:12513, PI: William Blair) will expand the WFC3 dataset available
for M83 from 2 to 7 fields.

M83 (NGC\,5236), also known as the ``Southern Pinwheel'' galaxy, is a
slightly barred spiral galaxy with a starbursting nucleus located at a
distance of $4.61$\,Mpc, i.e., $(m-M)_{0} = 28.32 \pm 0.13$
\citep{saha06}. The \ha emission can be used to pinpoint regions of
recently formed massive stars along the spiral arms, while red
supergiants can be found throughout the galaxy. The color-magnitude
diagram (CMD) of resolved stars is a powerful diagnostic tool for
understanding the stellar evolution and history of star formation of
galaxies in detail. By comparing stellar evolution models to observed
CMDs, we are able to determine the ages of the stellar populations in
galaxies. However, in active star-forming regions, the stars are often
partially obscured by dust. Applying a single extinction correction
value for the whole galaxy often results in over- or under-estimates
of the ages of individual stars.  If the CMD is the only tool used to
determine the ages, the spatial variations of dust extinction in
galaxies and their effects on the determined ages are not readily
apparent. The additional information available from the color-color
diagram can remedy this problem.  By using techniques developed in
this paper, we can constrain the variations of dust extinction across
M83, and make corrections for individual stars.  We also focus on
spatial variations in the stellar properties throughout the galaxy.

The \hst\/ WFC3 observations of M83 were performed in August 2009 as
part of the WFC3 Science Oversight Committee (SOC) Early Release
Science (ERS) program (ID:11360, PI: Robert O'Connell). The central
region (3.2\,kpc$\times$3.2\,kpc) of M83 was observed in August, 2009.
A second adjacent field to the NNW was observed in March 2010, and
will be included in future publications. Details of the WFC3 ERS data
calibration and processing are given by \citet{chandar10} and
\citet{windhorst11}.

In this paper we present \ubvi\/ photometry of resolved stars in M83
and the resulting color-color and color-magnitude diagrams.  We use
$(F336W\!-\!F438W)$ vs. $(F555W\!-\!F814W)$ color-color diagrams to
constrain the variation of dust extinction along the line-of-sight of
each individual star, and correct $F814W$ vs. $(F555W\!-\!F814W)$ CMDs
for the extinction of individual stars to determine their ages. The
high sensitivity and the superb resolving power in the WFC3 F336W-band
plays a key role in allowing us to develop our extinction correction
techniques, and demonstrate the performance of the newly installed
WFC3. This results in improved stellar age estimates, and a better
understanding of the recent star-formation history of M83.

Our investigations are focused on the followings:
\vspace{-0.5cm}
\begin{enumerate}[(1)]
\item Do we see spatial variations of stellar ages in M83? If so, can
  we use these variations to learn more about the evolution of the
  galaxy and what triggers star formation?
\vspace{-0.2cm}
\item How well do various age estimates correlate (i.e., resolved
  stars, integrated light, clusters, number ratio of red-to-blue
  stars, \ha\/morphology, stellar surface brightness fluctuations,
  presence of Wolf-Rayet stars)?
\end{enumerate}

This paper is organized as follows.  In \S 2, we describe the
observations, photometric analysis, and the extinction correction
method. CMDs and color-color diagrams of the 50~selected regions are
then used to provide stellar age estimates, as described in \S
3. Comparisons with other age estimates (i.e., integrated light and
star clusters) are also made.  In \S 4, we compare the stellar ages to
a variety of other parameters that correlate with age, including
red-to-blue star ratios, $H\alpha$ morphology, and surface brightness
fluctuations. Section 5 includes a discussion of how these comparisons
can be used as diagnostics, with special attention paid to the
question of what spatial variations can tell us about the star
formation history in M83, and on the question of how star clusters
might dissolve and populate the field.  A summary of our primary
results is provided in \S 6. In Appendix A, we investigate the
locations of sources of Wolf-Rayet stars in our M83 field and discuss
the correlation with the positions of young regions. In Appendix B, we
describe how the spatial resolution affects our measured colors and
magnitudes, and our age estimates of stars in our M83 field.

%                                       

%
%--------------------------------------------------
%%%%%%%%%% S2 Observation and Reduction %%%%%%%%%%               
\section{HST WFC3 Observations}
\subsection{The WFC3/UVIS Data}
The data were obtained in seven broad- and seven narrow-band filters
in the WFC3 UVIS and IR channels. Details are described in
\citet{dopita10} and \citet{chandar10}. In the current paper we use
four broad-band images. The filters and exposure times are F336W (1890
sec), F438W (1880 sec), F555W (1203 sec), and F814W (1203
sec). Although we do not make transformations to the
Johnson-Kron-Cousins \ubvi\/ system, we will refer to these filters as
\uu, \bb, \vv, and \iw\/ for convenience. Three exposures were taken
in each filter at different dithered positions to remove cosmic rays,
the intrachip gap, and to partially compensate for the undersampled
point-spread function (PSF) of the WFC3/UVIS channel.
The raw data were processed using the {\tt MULTIDRIZZLE} software
\citep{koekemoer02} with an effective pixel size of 0$\farcs$0396,
which corresponds to 0.885 pc per pixel at the distance of 4.61\,Mpc
\citep{saha06}. The resulting multidrizzled images are combined,
cosmic-ray removed, and distortion-corrected.  A color composite of
the F336W, F555W, F814W (broad bands) and F502N, F657N (narrow bands)
images is shown in Figure \ref{fig:m83_reg50}. It covers the nuclear
region of M83, part of its eastern spiral arm, and the inter-arm
region. Stars in the cores of compact star clusters are not resolved
at this resolution, but the brighter stars in the outskirts of the
star clusters and in the field are
generally resolved into individual stars. Details of the effects of
spatial resolution are discussed in Appendix B.  Since we are
interested in spatial variations of stellar ages in M83, we selected
50~regions in the spiral arm and inter-arm areas.
Boxes outlined in blue (very young -- age\,$\lesssim$\,10\,Myr, see \S
3), yellow (young -- 10\,$\lesssim$\,age\,$\lesssim$\,20\,Myr), and red
(intermediate-aged -- age\,$\gtrsim$\,20\,Myr) in Figure
\ref{fig:m83_reg50} show the 50~regions selected for detailed study in
this paper.

\subsection{Photometry and Artificial Star Tests}
Photometric analysis of the WFC3 M83 data was performed on the \uu,
\bb, \vv, and \iw\/ images using the {\tt DoPHOT} package
\citep{schechter93} with modifications made by A.~Saha.  Additional
routines to derive parameters for the kurtosis of the analytic PSFs
used by {\tt DoPHOT}, and for post-processing {\tt DoPHOT} output to
obtain calibrated aperture-corrected photometry, were performed using
customized {\tt IDL} code written by Saha. A more detailed description
of these procedures can be found in \citet{saha10}. The kurtosis terms
were derived by fitting the functional form of the {\tt DoPHOT}
analytic PSF to 20 relatively bright and isolated stars. The remaining
shape parameters ($\sigma_{x}$, $\sigma_{y}$, and $\sigma_{xy}$) are
dynamically optimized within {\tt DoPHOT}.

The difference between the shape parameters for an individual object
and those for a typical star was calculated and used to classify the
object as a star, a galaxy, or a double-star. The shape parameters
were determined separately for each image. The numbers of objects
classified as stars are approximately 20,000 in \uu\,and \bb, 17,000
in \vv, and 15,000 in \iw\,band images, respectively. The calculated
Full Width at Half Maximum (FWHMs) are \uu\,=\,0$\farcs$096 (2.42
pixel), \bb\,=\,0$\farcs$093 (2.34 pixel), \vv\,=\,0$\farcs$091 (2.29
pixel), and \iw\,=\,0$\farcs$113 (2.85 pixel). All magnitudes are in
the WFC3-UVIS VEGAMAG magnitude system, calculated using equation (4)
in \citet{sirianni05} and the latest zeropoint magnitudes:
\uu\,=\,23.46, \bb\,=\,24.98, \vv\,=\,25.81, and \iw\,=\,24.67 mag,
provided by STScI at the \hst/WFC3
website\footnote{http://www.stsci.edu/hst/wfc3/}.

A multi-wavelength \ubvi\/ catalog was then constructed by matching
the individual objects in each catalog between these filters.
 %using {\tt
%  TOPCAT\footnote{TOPCAT: Tool for OPerations on Catalogues And Tables
%    (http://www.star.bristol.ac.uk/$\sim$mbt/topcat).}} 
Interestingly, we found that the extreme image crowding in these
images, combined with the excellent panchromatic sensitivity and
unprecedented spatial resolution provided by the newly installed WFC3,
introduced a new challenge during this matching step, with large
numbers of ``pseudo'' matches occurring if too-large a match radius
was used. We discuss the effects of spatial resolution in Appendix B.

Figure \ref{fig:pseudo} demonstrates this problem by showing the
$F555W$ and color composite image cut-outs near Regions \#\,26 and
\#\,27. In the middle panel of Figure \ref{fig:pseudo}, stars circled
in red are objects with \vi\/ colors that are very red
($(V-I)>1.2$~mag), but have \uub\/ values that are fairly blue
($-0.4<(U-B)<-1.7$~mag). These can be seen as the spray of points in
the upper right parts of the top right panels of Figures
\ref{fig:allstar_no} and \ref{fig:allstar_cor}, which will be
discussed below. As shown by the postage stamp images in Figure
\ref{fig:pseudo}, about half of these are likely to be blue stars with
very high values of reddening (i.e., the bottom panel and the small
postage stamps on the right show that they are in dusty regions),
while the other half are very close superpositions of at least one red
and one blue star, which we will call ``pseudo'' matches (i.e., the
panels on the left show a strong blue and red gradient across many of
the objects).

We found that using a very small matching radius of 0.5 pixels was
needed to minimize ``pseudo'' matches in our catalog. Earlier attempts
using a matching radius of 3 pixels resulted in $\sim$10~\% pseudo
matches.  Accurate matching also requires precise geometric distortion
corrections in all filters.  However, even when these conditions are
met, there is a finite chance that two different stars will fall
within the same aperture. This is most easily seen by blinking the \uu-
and \iw-band images. While the single stars stay in the same position,
the pseudo match stars move slightly, showing they are two
different stars with different colors.

The final stellar catalog contains $\sim$15,000 objects. We find that
our procedure results in only a few percent of pseudo matches, as will
be discussed in \S 2.4 (i.e., the upper right panel of Figure
\ref{fig:allstar_cor}).

To measure the photometric completeness, we performed artificial star
tests with the same detection and photometry procedure as applied to
the actual stars. We inserted 1000~artificial stars with Gaussian PSFs
and FWHMs appropriate for these stars at random positions into all
four images. The magnitude of the inserted stars varied from 22 to 28
mag in steps of 0.25 mag.  The 50\% completeness levels at 5~$\sigma$
detection thresholds for a typical region are reached at recovery
magnitudes of approximately
24.3, 25.1, 25.3, and 24.8 mag for the \uu, \bb, \vv, and \iw\/
images, respectively. These limits can vary by about a magnitude,
depending on the brightness of the background in a given region.  The
exception is Region \#\,48, which includes the nucleus. The completeness
thresholds are two or more magnitudes brighter in this region due to
the very high background. This region has therefore been excluded from
the rest of the discussion, but is included in Table \ref{tab:50reg}
for reference.

\subsection{CMDs and Color-color Diagrams in M83}
The top panels of Figure \ref{fig:allstar_no} show the \iw~vs. ($V-I$)
CMD ({\it left}) and the \uub\, vs. \vi\, color-color diagram ({\it
  right}) of resolved stars extracted from the cross-matched catalog.
The CMD shows the presence of young main-sequence (MS) stars,
transition or He-burning ``blue loop'' stars, red giant stars
(hydrogen shell-burning and hydrogen$+$helium shell-burning stars),
and a relatively small number of low-mass red-giant branch (RGB)
stars.  This is because the tip of the low-mass RGB stars is at \iw\,=
24.7 mag \citep{karach07}, which is roughly the same as our
completeness threshold. The ages of these stars range from $\approx$ 1
to $\approx$ 100 Myr. The Padova isochrones \citep{marigo08} ranging
in age from log~$\tau$ (age/yr) = 6.55 (3.5\,Myr) to log~$\tau$
(age/yr) = 8.0 (100 Myr) are included in Figure \ref{fig:allstar_no}
for reference.  We note that the younger isochrones (e.g., 1\,Myr) fall
nearly on top of the 3.5~\,Myr isochrone, and hence are not included in
the diagram.

The arrows in both panels show $A_{V}\!=\!1$~reddening vectors for M83
using the $R_{V}=3.1$ extinction curve of
\citet{cardelli89}. Corrections have been made for foreground Galactic
extinction \citep{schlegel98} using $A_{F336W}=0.361$,
$A_{F438W}=0.290$, $A_{F555W}=0.229$, and $A_{F814W}=0.133$ mag,
respectively.\footnote{NASA/IPAC Extragalactic Database (NED):
  http://ned.ipac.caltech.edu/} In Figure \ref{fig:allstar_no}, we
only plot stars with photometric errors less than 0.25 mag in the
filters used for each diagram. The numbers of objects plotted in the
CMD and the color-color diagram are about 12,000 and 8,500 stars,
respectively. Since approximately 30\%~of the stars are not detected
in one or more of the \uu, \bb, or \iw\/ filters, due to either
reddening or intrinsically red or blue colors, fewer stars are plotted
in the color-color diagram.

We adopt a value of 1.5 times solar metallicity (Z=0.03), the highest
metallicity isochrones available from the Padova database, for M83
based on
\citet{bresolin02}. A test using solar metallicity (Z=0.019)
isochrones showed that the ages for a typical region would be
$\approx$ 2 Myr older if the lower metallicity is adopted.
This agrees with the study by \citet{larsen11}, who found that the
simulated CMDs of young massive clusters in M83 with solar and
super-solar metallicity isochrones would not look much different.
Theoretical isochrones calculated for the WFC3 filters and Z=0.03 (1.5
Z$_{\sun}$) from the Padova database\footnote{CMD version 2.3;
  http://stev.oapd.inaf.it/cgi-bin/cmd} \citep{marigo08} are overlaid
on both panels in Figure \ref{fig:allstar_no}.  The dashed line in
cyan shows the 50\%~completeness level in \iw\/ and \vi, using the
completeness threshold numbers from the artificial star test in \S
2.2.

The bottom panels of Figures \ref{fig:allstar_no} and
\ref{fig:allstar_cor} show the histograms of ages and masses for the
same stars plotted in the CMD and color-color diagram. The ages and
masses of stars were estimated by finding the closest match between
the \iw\/ and \vi\/ values for each star, using a fine mesh of the
stellar isochrones plotted in the CMD and color-color diagram in
Figures \ref{fig:allstar_no} and \ref{fig:allstar_cor}. More details
about the stellar age-dating from the CMDs and the stellar isochrones
are given in \S 3.1.

\subsection{Extinction Corrections}
Since M83 has an intricate structure of dust lanes associated with
active star forming regions in the spiral arms and thin layers of dust
in the inter-arm area in Figure \ref{fig:m83_reg50}, we cannot use a
single value of internal extinction and apply it to correct for the
extinction of all stars in a given region.  In the color-color diagram
in Figure \ref{fig:allstar_no}, we notice that while the bluest stars
match the model isochrones quite well after we correct for Galactic
foreground extinction, the majority of stars are found redward of the
models, implying a typical extinction of $A_{V}\!\simeq\!0.5$~mag (for
the densest part of the data ``swarm''). As described below, we can
determine the reddening of these stars {\it if} we assume that they
are intrinsically blue stars that belong on the Padova models (i.e.,
with $(V-I) \simeq -0.3$~mag).  A visual inspection supports this
interpretation. Stars with observed values of $(V-I) \approx -0.3$~mag
are found in areas with no obvious dust features surrounding them
while stars with $(V-I) > 0.5$~mag are near dust filaments.

Our basic approach to correct the colors (and luminosities) of
individual stars for the effects of extinction is to backtrack the
position along the reddening vector for each data point, until we hit
the stellar isochrones in the color-color diagram. What makes this
method work is the fact that all the isochrones are nearly on top of
each other in the range $(U\!-\!B)\!<\!0.0$~mag.  The way we apply this in
practice is to match the observed and predicted values of the
reddening free $Q$ parameter defined by:
\begin{equation}
Q_{UBVI} = (U-B) - \frac{E(U-B)}{E(V-I)} \times (V-I)
\end{equation}
\citep{ga81}. Using observed values of \ubvi, we
%find the $(U-I)$ color of each star and 
compute the observed $Q_{UBVI}$ values by using the standard slope of
$E(U\!-\!B)/E(V\!-\!I)=0.58$ \citep{ga81,whitmore99}, which
corresponds to $R_V = 3.1$.
%of stars with the ratio of 0.58 of $(U-B)$ to $(V-I)$ color excesses. 
The predicted $Q$ values are calculated using the 1.5~Z$_{\odot}$ Padova
isochrones.
A complication is that we cannot simply use the matched $Q$ values to
determine corrected values of \uub\/ and \vi, since they would,
by definition, fall precisely on the isochrones (i.e., we would be
using circular reasoning). Instead, we solve for the extinction in
$(U\!-\!I)$, which is partly independent and uses the longest wavelength
baseline. We then calculate the extinction values in \ubvi\/ and the
color excess values in \uub\/ and \vi.
The averages of the corrected internal extinction and color excess
values by our star-by-star correction method are $A_{F336W}=$ 0.696,
$A_{F438W}=$ 0.559, $A_{F555W}=$ 0.441, $A_{F814W}=$ 0.256 mag, and
$E(U\!-\!B)=$ 0.137 and $E(V\!-\!I)=$ 0.185 mag. This is similar to
the average internal extinction ($A_{F555W}=0.425$ mag) estimated from
the cluster SED fitting by \citet{bastian11}. However, our results
differ from the average extinction ($A_{V}=0.671$\,mag) of 45 clusters
in the nucleus ($\sim20\arcsec$ in diameter) of M83, determined from
$H\alpha/H\beta$ ratios by \citet{harris01}. This is consistent with
the finding of larger extinction in the nuclear region by
\citet{chandar10}.

In the color-color diagram (top right panel) of Figure
\ref{fig:allstar_no}, we note that while most of the data points are
consistent with the standard reddening vector, the stars located in
the green triangle appear to follow a flatter reddening
vector. Whether this is actually due to a different reddening law in
M83 (e.g., as has been suggested for heavily extincted regions such as
\#\,4, \#\,12, and \#\,20), or is some sort of artifact (e.g. due to
photometric uncertainties, unresolved star clusters, or the ``pseudo''
matching problem discussed above) is difficult to determine from our
present dataset.  However, for our specific needs the precise answer
to this question is not critical.  Our approach will be to correct the
data points in the triangle back to a position near the top of the
isochrones, as shown by the intersection of the red dotted line and
the isochrones in Figure \ref{fig:allstar_cor}. This is equivalent to
using a range in $R_{V}$ between 3.1 and 5.7 (the value represented by
the red dotted line).

The upper panels of Figure \ref{fig:allstar_cor} show the CMD and
color-color diagram corrected for the internal extinction in M83. The
arrows in both panels are the reddening vectors.  We do not apply an
extinction correction for stars with $(V-I)\geq1.2$~mag, or above an
extrapolation of the flatter extinction vector from the bluest
possible isochrones (see the red dotted lines in the upper right panel
of Figure \ref{fig:allstar_cor}), since we believe that the colors of
many of these sources are incorrect (i.e., roughly half are likely to
be pseudo matches, as discussed in \S 2.2 and shown in Figure
\ref{fig:pseudo}).  We note that only about 4\,\% of the stars fall in
this part of the diagram. These stars are removed from the subsequent
analysis. If we include these stars, our age estimates typically
increase by about 1\,Myr.
 
Also, stars below the blue dotted line in Figure \ref{fig:allstar_cor}
are not corrected for extinction, since these are main sequence
turnoff stars (i.e., blue loop or transition stars in general), and
hence, do not have unique $UBVI$ colors for the stars along a given
reddening line, unlike the main sequence stars with $(U\!-\!B)\!<\!0.0$~mag.

Based on the extinction corrected CMD in Figure
\ref{fig:allstar_cor}, we note that: (1) the corrected data now shows
a relatively narrow distribution of points along the left side of the
CMD in agreement with ages of $\sim$3 Myr, and
(2) while the data now aligns with the model isochrones much better,
there is still a fair amount of scatter (especially for $I$~$<$~24),
primarily due to observational uncertainties from the fainter stars.

The bottom panels of Figure \ref{fig:allstar_cor} show the distribution
of ages and masses of stars after we apply our extinction correction
technique. 
Based on the automatic method described in \S 3.1, we note that the
luminosity weighted mean ages determined from the corrected colors and
luminosities are significantly younger
($\sim$14~Myr; Figure \ref{fig:allstar_cor}) than the mean ages
determined from the uncorrected colors and luminosities ($\sim$27~Myr;
Figure \ref{fig:allstar_no}), as we expected.

%
%--------------------------------------------------
%%%%%%%%%% S3 Resolved Stars %%%%%%%%%%                            
%\section{Results and Discussion}
\section {Age-Dating Populations in M83}

In this section we discuss several different methods of age-dating
stellar populations in M83, using both individual stars and star
clusters, and then intercompare the results.
We do not necessarily expect field star and cluster ages to agree,
since the dissolution of clusters may bias the observed cluster
population towards younger ages than the surrounding field stars.  

\subsection{Age-Dating the Resolved Stellar Population using Color Magnitude Diagrams}
As shown in Figure \ref{fig:m83_reg50}, we selected 50~regions that
cover spiral arm and inter-arm areas in order to study the recent
star-formation history of M83. 
Figures \ref{fig:reg12_7} and \ref{fig:reg29_2} show
cutouts of Regions \#\,12, \#\,7, \#\,29, and \#\,2.\footnote{The
  color-magnitude diagrams and color-color diagrams of all 50 regions
  (similar to Figures 5 and 6) are available on the WFC3 Early Release
  Science Archive website (http://archive.stsci.edu/prepds/wfc3ers/).}
These figures include color-color diagrams of the stars in these
regions, as well as CMDs that are uncorrected (upper) and corrected
(lower) for extinction.  These four regions highlight the range in age
of the dominant stellar population, from very young to intermediate
ages: i.e., Region \#\,12 with very strong \ha emission superposed on
the cluster stars; Region \#\,7 with a large bubble of \ha emission
surrounding the stars and clusters; Region \#\,29 with no \ha
emission, but large numbers of bright red and blue stars; Region \#\,2
with no \ha emission and fainter stars.

A cursory glance at the CMDs of these four regions (Figures
\ref{fig:reg12_7} and \ref{fig:reg29_2}) show clear differences. The
primary differences are, as predicted by the isochrones, that younger
regions contain: (1) bluer main sequence stars; (2) larger numbers of
upper main sequence stars; and (3) larger ratios of blue to red
stars.  Our regions do not, however, contain only stars of a single
age.  Even these four regions, which were chosen to include a single
dominant stellar population, contain a mix of young and old stars.  In
this paper, we are primarily interested in the ages of the bright
young stars in these regions, which dominate CMDs in
luminosity-limited samples.

In practice, this focus on the younger population is carried out by
imposing a magnitude cutoff of $M_I=-5.5$~mag (i.e., $I=22.82$
mag which corresponds to the age cutoff of $\sim$60\,Myr in the CMD)
for the stars used to estimate ages (i.e., the magenta dashed line in
Figure \ref{fig:allstar_cor}).  This luminosity limit allows us to
stay above the completeness limit for all but the very reddest stars
(see Figure \ref{fig:allstar_no}), while also focusing on evolved
stars and those on the upper main sequence, which are most sensitive
to the age of a stellar population.  Our primary goal is to obtain
reliable ``relative'' estimates for the 50~regions (rather than
absolute ages), and our magnitude limit is sufficiently deep to
accomplish this goal.

We compare the extinction corrected colors and magnitudes of stars
with the Padova stellar isochrones in two ways to estimate the age of
the dominant stellar population in each region.  First, two of the
authors (HK and BCW) independently estimated ages based on a visual
comparison of the CMDs and isochrones. We focused on features such as
the color of the main sequence stars, the number of stars in the upper
main sequence, and the number ratio of blue to red stars.  The left panel
of Figure \ref{fig:cmdages} shows that the independent, visually
determined age estimates are in good agreement with a 7.9~$\sigma$
(slope/uncertainty of the best linear fit) correlation and a slope
within 1~$\sigma$ of the unity value.
The average ages of these two independent (manual) estimates are listed
in column $Age_{CMD}$$_{man}$ in Table \ref{tab:50reg}.

Next, we estimated the ages automatically, finding the closest match
in both age and mass for each star by comparing the extinction
corrected \iw~magnitude and \vi~color with a fine mesh of stellar
isochrones generated from the Padova models (log$\tau$ (age/yr)
ranging from 6.05 to 8.35 in steps of 0.05).  We then calculate a
luminosity-weighted age from the distribution of individual stellar
ages in each region. The result of the automatic age estimate is
listed in column $Age_{CMD}$$_{auto}$ in Table \ref{tab:50reg}.

To assess the uncertainties in our age and mass determinations for
each star, we ran a test for four different cases by adding and
subtracting the photometric errors in the \vi~color and \iw-band
magnitude. This moves each object on the CMD to the right
($+(V\!-\!I)_{error}$), left ($-(V\!-\!I)_{error}$), down ($+m_{I_{error}}$),
and up ($-m_{I_{error}}$). We then estimated the ages for these four
cases in the same way as we did to get $Age_{CMD}$$_{auto}$. The
results of this test are summarized in column
$\Delta$$Age_{CMD}$$_{auto}$ in Table \ref{tab:50reg}.
As shown in the $\Delta$$Age_{CMD}$$_{auto}$ column in Table
\ref{tab:50reg}, the variation in the age estimates are typically only
a few Myr (i.e., $\lesssim$10\%).

Figure \ref{fig:age_mass} shows the age and mass histograms of
individual stars in the four regions shown in Figure \ref{fig:reg12_7}
and \ref{fig:reg29_2}.  Luminosity-weighted mean values of the stellar
ages are shown in the upper right of the left panels. The age sequence
of these regions is as expected, from the youngest in Region \#\,12 to
the oldest in Region \#\,2.

One unexpected result was that while Region \#\,2 was selected to be
representative of an older field population, it also contains a fair
number of bright young stars. This suggests that young individual
stars may form in the field in low numbers. Another possibility is
that many of these are ``runaway'' stars, i.e., young stars that have
been dynamically ejected from their birth-sites within young star
clusters due to interactions with other stars.  \citet{whitmore11}
also suggested this as a possible explanation for isolated HII regions
ionized by a single, massive star, i.e., ``Single-Star'' HII (SSHII)
regions (see Figure 3 from Whitmore 2011).  Either
explanation would be interesting, and should be investigated in a
future study.  Figure \ref{fig:reg9} shows a similar ``field'' Region
\#\,9 with primarily old red
giant stars (barely discernible), but which also includes a small
population of isolated young blue stars and four compact clusters
ranging in age from intermediate to old.

The right panel of Figure \ref{fig:cmdages} compares the
$Age_{CMD}$$_{man}$ and $Age_{CMD}$$_{auto}$ estimates for the
different regions.  
A unity line is included for comparison, and shows that there is good
agreement (7.6~$\sigma$ correlation) between the two methods which
varies from the unity line by only 1~$\sigma$. We use the objective
luminosity-weighted age estimates ($Age_{CMD}$$_{auto}$) as our
preferred method for estimating ages of the resolved stellar
populations in the remainder of this paper.

%\subsection{The Pros and Cons of our Approach Compared to Other Studies}
\subsection{Comparison With Previous Age Estimates}

\citet{larsen11} recently estimated the ages of two star clusters in
M83 by comparing photometry of individual stars in the outskirts of
the clusters with isochrones.  These clusters are within Regions \#\,3
and \#\,11 here.  \citet{larsen11} used a more sophisticated method,
comparing observed and predicted magnitude distributions for red and
blue stars using Hess diagrams.  They did not however, correct the
magnitudes or colors of their stars for the effects of spatial
variation of extinction, but since these are intermediate age clusters
with minimal internal extinction, this is not a large effect in these
particular cases.  They find log $\tau$ of 7.44 (27.5~Myr) and 7.45
(28.2~Myr) for the clusters in Regions \#\,3 and \#\,11.  We find
similar ages of 21.9~Myr (Region \#\,3) and 26.7~Myr (Region \#\,11)
for the dominant luminous stellar populations, despite the fact that
our photometry includes a significantly larger area around each
cluster. Hence we find good agreement between our age estimates and
those of \citet{larsen11}.

Each method has its own strengths and weaknesses.  Our approach has
the advantage of correcting the photometry of individual stars for the
effects of extinction to improve our age estimates, but the
disadvantage that this requires imaging in more than three optical
bands with a crucial need for the \uu-band observation.
% to achieve a wide wavelength coverage as used in this study.
The \citet{larsen11} method does not correct for the effects of
extinction, but allows for a better determination of the star
formation history over a wider age range (i.e., out to $\sim$1\,Gyr).
% for the stellar population.

%\subsection{Age-Dating Clusters Using the Integrated Cluster Light and SED Fitting}
\subsection{Comparison between \boldmath{$Age_{CMD}$$_{auto}$} and Compact Cluster Ages}

Much of our recent work has involved age-dating star clusters using
the integrated light (\ubvi\/ and \ha emission) from the cluster and
model spectral energy distribution (SED) fitting. To compare the age
of star clusters in each region to the age determined from individual
stars in each region, we adopt the ages and luminosities of compact
clusters identified in our previous work \citep{chandar10}. Details of
the cluster age-dating method can be found in that paper.

We note that the regions sampled in this study are large (i.e.,
several hundred\,pc$^2$) and may include stars and star clusters
spanning a wide range of ages. This requires making the comparison
between the ages of field stars and star clusters with caution. Only
the largest star forming complexes within spiral galaxies have such
dimensions (e.g., the giant HII region NGC\,604 in M33;
Freedman et al. 2001). More typical regions are much smaller, and hence
would not dominate the entire field. This will tend to weaken our
correlations, especially when comparing resolved stellar ages with
cluster ages.

Using the ages of star clusters determined from the SED fitting, we
compute the luminosity-weighted mean age of each region, as we did for
stars in the same region. The ages of 50~regions are listed in column
$Age_{Cl}$ in Table \ref{tab:50reg}. The upper panel of Figure
\ref{fig:sedages} shows a comparison between the luminosity-weighted
mean age ($Age_{Cl}$) of the compact clusters in a given region and
our stellar age ($Age_{CMD}$$_{auto}$) estimates based on the CMDs of
all the resolved stars in the region (as discussed in \S 3.1). The
correlation between these two ages is fair (5.4~$\sigma$
correlation). We note that while the midpoints for the two methods are
similar, the slope is steeper, as discussed in more details in the
next section.

%\subsection{Age-Dating Regions Based on Integrated Colors}
\subsection{Comparison between \boldmath{$Age_{CMD}$$_{auto}$} and Age Estimates from Integrated Light from the Entire Regions}

We obtain an independent estimate of the age of each region by
measuring the colors of the entire region, and perform a simple SED
fit, in the same way as we did for star clusters
\citep[i.e.,][]{chandar10}.  The lower panel of Figure
\ref{fig:sedages} shows a comparison between our age estimates for the
resolved stellar populations using CMDs ($Age_{CMD}$$_{auto}$) and
from the integrated light ($Age_{Reg}$) in each region.

While there is a fair correlation (4.6~$\sigma$) between the CMD and
integrated photometric age estimates, there is also a fair amount of
scatter and an apparent gap in the region ages in the range 6.8 $<$
log (age/yr) $<$ 7.2. This gap is similar to the well known artifact for
cluster age estimates which is due to the looping of predicted cluster
colors in this age range (see Chandar et al. 2010).  We also note that
the slope of the relation is steeper than the unity vector, with
integrated light ages ranging to much lower ages than the CMD
ages. This is similar to the comparison with $Age_{Cl}$, as discussed
above. This is probably caused by a variety of effects including: 1)
the integrated age estimates take into account \ha emission (which is
very sensitive to massive young stars) while the CMD estimates do not;
2) the 1~Myr isochrones are essentially on top of the 3~Myr
isochrones, making it rare that the minimum age ever gets selected by
the software that does the matching with the isochrones; and 3) the
adoption of a $M_V = -5.5$~mag cutoff removes all stars with ages
$\gtrsim$100 Myr from the $Age_{CMD}$$_{auto}$ estimates,
while the light from older stars is still included in the integrated
light used for the $Age_{Reg}$ estimate (hence, the
$Age_{CMD}$$_{auto}$ estimate can never be $\gtrsim$50~Myr, while the
$Age_{Reg}$ can be older).
% a large number of young blue stars but a smaller number of old red
% stars.

This difference in slope in Figure \ref{fig:sedages} highlights the
fact that each age-dating method has its own idiosyncrasies. By
comparing a number of different methods, we begin to understand these
artifacts better, and learn how large the true systematic uncertainties can
be.

%
%--------------------------------------------------
%%%%%%%%%% S4 Discussion %%%%%%%%%%
%\section{Discussion}
\section{Comparison between \boldmath{$Age_{CMD}$$_{auto}$} and Other Parameters that Correlate with Age}

We are now ready to compare our age estimates for the resolved stellar
components in 50~regions to other observables that correlate with
age. These include: (1) number ratio of red-to-blue stas
\citep[e.g.,][]{larsen11}; (2) morphological categories
\citep{whitmore11};
%(3) average extinction of individual stars in each region 
and (3) stellar surface brightness fluctuations \citep{whitmore11}. 
%We will discuss each of these in their section.

%
%
\subsection{Comparison between \boldmath{$Age_{CMD}$$_{auto}$} and Number Ratio of Red-to-Blue Stars}
One property that is expected to correlate with age is the number
ratio of red-to-blue stars. At very young age, all stars are blue. As
the population ages, the number of red giant stars (H-shell burning
and H$+$He-shell burning stars) increases. This effect can be seen by
noting that nearly all of the stars in Region \#\,12 (Figure
\ref{fig:reg12_7}) are blue, while there are large numbers of red
stars in Regions \#\,29 and \#\,2 (Figure \ref{fig:reg29_2}).

The number ratio of red-to-blue stars (column Red-to-Blue Ratio in
Table \ref{tab:50reg}) is calculated
%the age-dating from the isochrone fitting. Stars 
by using a criteria that \vi\/ be redder than 0.8 mag for red
stars. The top panel of Figure \ref{fig:rb_ha_rms} shows a fairly good
correlation (7.2~$\sigma$) between $Age_{CMD}$$_{auto}$ and
the Red-to-Blue Ratio.
% except for regions with low statistics ($N_{star} < 15$).

%
\subsection{Comparison between \boldmath{$Age_{CMD}$$_{auto}$} and Morphological Category}
The middle panel of Figure \ref{fig:rb_ha_rms} shows the correlation
between our automatic CMD age estimates and the morphological
classification for a given region, as defined in \citet{whitmore11}.
Briefly, regions with \ha emission superposed on top of the stellar
component are type 3 (emerging star clusters), regions with small \ha
bubbles are category 4a (very young), regions with large \ha bubbles
are category 4b (young) , and regions with no \ha bubbles are category
5 (intermediate age). Based on this criteria, the regions displayed in
Figures \ref{fig:reg12_7} and \ref{fig:reg29_2} are classified as
categories 4a (Region \#\,12), 4b (Region \#\,7), 5a (Region \#\,29), and
5a/5b (Region \#\,2), respectively.
Each region was classified independently by two authors (HK and BCW).
The mean of the two determinations is used in what follows, and is
listed in column ``$H\alpha$ Morphology'' of Table \ref{tab:50reg}.

The middle panel of Figure \ref{fig:rb_ha_rms} shows that stellar CMD
age estimates ($Age_{CMD}$$_{auto}$) and morphological categories
(\ha\/morphology) show a fair correlation (5.8~$\sigma$), but
not as good as the strong correlations (9~$\sigma$ for log~$\tau<7$ and
5~$\sigma$ for log~$\tau>7$) found with ages for compact clusters in
\citet{whitmore11}.  This is probably due to the fact that most of the
regions are not dominated by a single age stellar population, making the
morphological classification for an entire region problematic (i.e.,
there is strong \ha emission in part of the region but none in other
parts).

\subsection{Comparison between \boldmath{$Age_{CMD}$$_{auto}$} and Stellar Surface Brightness Fluctuations}
In \citet{whitmore11}, we developed a method to age-date star clusters
based on an observed relation between pixel-to-pixel flux variations
(RMS) within star clusters and their ages. This method relies on the
fact that the young clusters with bright stars have higher
pixel-to-pixel flux variations, while the old clusters with smoother
appearance have small variations.  This is because the brightest stars
in a 100~Myr population have $M_{I} \approx -3$ mag, which is below
our 50\,\% photometric completeness limit ($M_I=-3.6$\,mag) and well
below our magnitude cutoff of $M_{I}=-5.5$\,mag used in our stellar
age-dating method described in \S\,3.1.

The bottom panel of Figure \ref{fig:rb_ha_rms} shows the correlation
between the resolved stellar ages ($Age_{CMD}$$_{auto}$) and the
surface brightness fluctuations (Pixel-to-Pixel RMS) measured in the
50 selected regions. We find little or no correlation (1.1~$\sigma$)
with a relatively large amount of scatter, especially for the younger
clusters. Part of this scatter may be caused by the two-valued nature
of the effect as discussed in \citet{whitmore11}, with a maximum value
of the surface brightness fluctuations around 10~Myr, and lower values
for both smaller and larger ages. Another reason for the scatter is
the fact that many of the regions have mixed populations, rather than
single age populations.

%%% S5 Discussion  %%%%%%%%%%%
%\section{Discussion}
%
\section{Insights into the Star Formation History in M83 Based on Spatial Variations for 50~regions}
We first examine how the ages for the 50~regions are distributed in
Figure \ref{fig:m83_reg50}.  The boxes in this figure are color-coded
as follows: blue (very young with ages less than about 10 Myr), yellow
(young with ages between about 10 and 20 Myr), and red (intermediate
aged with ages greater than about 20 Myr). As expected, we find that
the very young regions are primarily associated with the spiral arms,
the young ryoungs tend to be ``downstream'' from the spiral structure
(i.e., on the clockwise side away from the dust lanes), while the
intermediate aged regions are found on both sides of the spiral arms
in the interarm regions.

A similar, more detailed analysis is being done by Chandar and Chien
using the cluster ages in M51 (2012, in preparation) and M83 (2012, in
preparation). They compare the age gradients they find with models
developed by \citet{dobbs10}, assuming a variety of different
triggering mechanisms (i.e., spiral arms, bars, stochasticity, and
tidal disturbances).

We can perform a similar experiment here, since we have age estimates
for each individual star.  The three panels in Figure
\ref{fig:pellerin} show the distribution of stars with ages in the
ranges 1--10, 15--35, and 40--100 Myr, respectively. The top panel is
the youngest group of stars, clearly showing that the stars in these
regions are mostly distributed along the active star-forming region
(i.e., associated with strong \ha emission) in the spiral arms.  
%This provides support for the idea that shock waves associated with
%the spiral arms are the dominant trigger of star formation.
The stars in the middle panel tend to be found slightly downstream of
the spiral arms, while the older stars are still farther out in the
inter-arm regions, as expected.

These diagrams can also be used to examine whether there is evidence
that most of the young stars form in clusters and clustered regions
and then dissolve into the field. \citet{pellerin07} made a similar
set of diagrams for the galaxy NGC\,1313, which supported this
interpretation.  We find that the young star samples show strong
clustering, while the older star samples are progressively more
uniform.  Hence, these distribution maps of stars in the different age
group support the idea that the stars form in clusters in spiral arms,
and then most of the clusters dissolve, populating the field with
stars \citep{lada03}. Chandar et al. (2012, in preparation) will
examine this subject more quantitatively in the future.

%%%%%%%%% S6 Summary %%%%%%%%%%%
\section{Summary}
Color-magnitude diagrams and color-color diagrams of resolved stars
from the multi-band \hst/WFC3 ERS observations of M83 have been used to
measure the ages of stellar populations in 50~regions of this well known
face-on
% starburst - BCW changed since starburst is manly in nuclear region
% and we mainly deal with outer galaxy.
spiral galaxy.  The diagrams show the presence of multiple stellar
features, including recently formed MS, He-burning blue-loop stars,
and shell-burning red giant stars.  Comparisons between our stellar age
estimates, and a wide variety of other age estimators, allow us to
investigate a number of interesting topics.  The primary new results
from this study are as follows:

(1) An innovative new technique using a combination of CMD and
color-color diagrams has been developed in order to correct for the
extinction towards each individual star and to age-date the stellar
population.  The mean extinction values for the 50~regions studied in
this galaxy are 0.696 ($A_{F336W}$), 0.559 ($A_{F438W}$), 0.441
($A_{F555W}$), 0.256 ($A_{F814W}$) mag, respectively.

(2) The various age estimators (stellar, integrated light, clusters
within the region) show fair correlations (i.e., between 4.6~$\sigma$
and 5.4~$\sigma$).  This is expected, since many of the regions have a
mixture of different-aged populations within them, and each technique
uses light from a different subset of the stars (e.g., the stellar age
estimates use only bright stars, while integrated light includes all
the light). A comparison with previous age estimates by
\citet{larsen11} shows a good agreement (see \S 3.2 for details).
% When only a single dominant population is seen within a - BCW
% commented out - not sure we looked at this and not sure if it is
% true region (eg., regions \#3, \#11, \#29, \#47) the agreement
% is quite good.

(3) Comparisons between the stellar ages and other parameters that are
known to correlate with age show a range from very good correlations
(e.g., with red-to-blue ratios: i.e., 7.2~$\sigma$) and fair
correlations (e.g., \ha morphology with 5.8~$\sigma$) to little or no
correlation
\citep[e.g., pixel-pixel RMS from][] {whitmore11}. This is expected
for reasons similar to the ones discussed above.

(4) The regions with ages younger than 10~Myr are generally
located along the active star-forming regions in the spiral arm.
%providing support for the idea that shock waves associated with the
%spiral arms are the dominant trigger of star formation in M83.
The intermediate age stars tend to be found ``downstream'' (i.e., on
the opposite side from the dust lane) of the spiral arms, as expected
based on density wave models.  A more detailed comparison with models
with various other triggering mechanisms \citep[e.g., bars, density
  waves, tidal disturbance, and stochasticity; see][]{dobbs10} is
in process by Chandar and Chien for M51 and M83 (2012, in
preparation).

(5) As discussed in the Appendix A, the locations of Wolf-Rayet
sources from \citet{hadfield05} are in broad agreement with the age
estimates discussed in the current paper. The much better spatial
resolution from \hst\/ shows that many of the Wolf-Rayet ``stars''
from ground-based observations are actually young star clusters.

(6) Effects of spatial resolution on the measured colors, magnitudes,
and the derived ages of stars in our M83 images are described in the
Appendix B. Based on a numerical experiment using a star cluster
NGC\,2108 in the LMC, we found that while individual stars can
occasionally show measurable differences in the colors and magnitudes,
the age estimates for entire regions are only slightly affected.\\

%% Acknowledgment
We thank Zolt Levay for making the color images used in Figure
\ref{fig:m83_reg50}. We also thank the anonymous referee for helpful
comments. This project is based on Early Release Science observations
made by the WFC3 Science Oversight Committee. We are grateful to the
Director of the Space Telescope Science Institute for awarding
Director's Discretionary time for this program. Support for program
\#\,11360 was provided by NASA through a grant from Space Telescope
Science Institute, which is operated by the Association of
Universities for Research in Astronomy, Inc., under NASA contract NAS
5-26555.

Facilities: \hst\,(WFC3)

%% Appendix
\appendix

%\section*{APPENDIX. Using Wolf-Rayet Stars to Age-Date Regions in M83}
\section*{Appendix A. Using Wolf-Rayet Stars to Check for Consistency with Our Age Estimates}

In this paper we compared the ages resulting from a variety of
age-dating methods for different star-forming regions in M83.  Here,
we extend our analysis by considering previously identified Wolf-Rayet
stars in M83, which are $\sim$1--4~Myr old \citep{crow07}, as another
independent age estimate.

\citet{hadfield05} identified 283 Wolf-Rayet sources in M83 from
narrow-band imaging centered on the He II $\lambda$4686 emission line,
plus spectroscopic follow-up.  Seventeen of these Wolf-Rayet
sources fall in our WFC3 field-of-view. Figure \ref{fig:wr1} shows
their locations in the \hst\/ image, and Table 2 lists the coordinates
of these sources, as well as the region in which they are located.

The lower resolution of the ground-based \citet{hadfield05} study
makes it difficult (in many cases) to uniquely identify the Wolf-Rayet
sources in the \hst\/ image, although we can easily determine whether
such a source is located within one of our 50~regions.  Figure
\ref{fig:wr1} shows a color image of M83 (center panel) with five
regions, labeled A through E, that contain all 17 Wolf-Rayet sources.
Here, the \ha and F814W images are shown in red, F555W in green, and
F336W in blue.  Black and white images are the image cutouts in the
F814W and the \ha filters of regions A--E, as outlined in green in the
central panel.  The original finding chart of the region A taken from
\citet{hadfield05} is shown in the top right panel in Figure
\ref{fig:wr1}.  In the \ha image, the stellar continuum has been
subtracted from the \ha image, leaving just the ionized gas.  We
matched the coordinates of the Wolf-Rayet sources and the \hst\/
images, by assuming that the Wolf-Rayet source \#\,99 is perfectly
centered on a relatively isolated, compact \ha knot. We also
  compared the locations of the 17 matched Wolf-Rayet sources in the
  M83 \hst\/ images with the finding charts from \citet{hadfield05} to
  verify our source matching.

The excellent spatial resolution of the \hst\/ image also makes it
possible to check whether any of the Wolf-Rayet sources are actually
compact star clusters. While the majority do appear to be individual
stars, or a dominant star with a little ``fuzz'' which is likely a
faint companion star, there are at least three cases where the
counterpart appears to be a bright compact cluster (i.e., \#\,74, 86,
105: circled objects in Figure \ref{fig:wr2}). These three objects are
brighter than $M_{V} = -9$ mag with the concentration index (CI)
larger than 2.3, consistent with the cluster CI used in
\citet{chandar10}. There are several more cases where the counterpart
is in a looser association of stars (e.g., \#\,78, 102, 103).
%The nature of object \#\,86 is particularly unclear, since its
%coordinates place it on the edge of a very bright compact cluster.

All the Wolf-Rayet sources, except the sources in cutout E and \#\,108
in cutout D, are located in the spiral arms. This is expected, since
these are the sites of most recent star formation.  Nine of the 17 are
in regions of strong \ha emission, seven more are in regions of faint
\ha emission, and only one source (\#\,107 in cutout D) appears to
have no \ha associated with it.

Our primary question is whether the Wolf-Rayet sources tend to be
found in regions that we estimated to have young ages.  In Table 2 we
match the Wolf-Rayet sources with the region where they are found in
and find that their ages range from 1.1~Myr (Region \#\,35) to
50.1~Myr (Region \#\,41). This shows that such a correlation does
exist, with 6/8 (75~\%) of Wolf-Rayet sources in regions having
$Age_{Reg}$$<$4.6~Myr.  Similarly, \citet{hadfield05} found that five
of their Wolf-Rayet sources are included in a compact star cluster
catalog developed by \citet{larsen04}, and three of these five
clusters (\#\,193, 179, 179) have estimated ages in the range 1.5 -- 6
Myr, consistent with ages expected for Wolf-Rayet stars.

Figure \ref{fig:wr2} shows the CMD and color-color diagram of the
measured colors from our \hst\/ observations of the object that best
matches the Wolf-Rayet sources from \citet{hadfield05}.  We find that
most of these sources are in the top left of the two-color diagram. We
conclude that there is a strong tendency for Wolf-Rayet sources to
be associated with the youngest regions of star formation. The good
correlation provides general support for the accuracy of our CMD age
estimates.

%%%%%
\section*{Appendix B. Effect of Spatial Resolution}

Spatial resolution affects virtually all studies of individual stars
in nearby galaxies.  As is typical, in this paper we have assumed that
the point sources identified in Section~2 are individual stars,
although we also discovered a small fraction of pairs of nearly
aligned stars, one red and one blue, that can be identified from their
anomalous colors (e.g., ``pseudo-match'' Figures 2 and 4).  Here, we
address more generally how spatial resolution affects the measured
colors and magnitudes, and hence the derived ages, of stars in our M83
regions.

We begin by answering a simple question: ``What would the Orion Nebula
look like at the distance of M83? Would we be able to distinguish the
four central stars that make up the Trapezium or would they appear as
a single star?''  The separation between the four stars in the
Trapezium ($10\arcsec$ at D\,$\approx0.5$\,kpc) would be roughly
$0.001\arcsec$ at the distance of M83, approximately a factor of
40~smaller than a single pixel in our image, and hence these would
appear as one point-like source.  However, we note that the vast
majority of the remaining bright stars in the Orion nebula are much
more widely separated, and would not suffer from this problem.  We
also note that our selection criteria are designed to avoid including
stars in the crowded central regions of star clusters and
associations, which also minimizes the problem.

We perform a more quantitative analysis by degrading $HST$/ACS images
of the compact star cluster NGC\,2108 in the Large Magellanic Cloud
(LMC) \citep[ID:10595; PI: Goudfrooij, see][]{goud11}, which is at a
distance of $\sim50$\,kpc, by a factor of 100 (i.e., equivalent with
the distance of $\approx5$\,Mpc), as roughly appropriate for M83.  The
degraded images, created from a combination of rebinning and Gaussian
smoothing to approximately mimic the point spread function of ACS, are
then run through our normal object-finding software described in
Section~2. The numbers of objects found in the degraded NGC\,2108
images are $\sim$20 in the F555W and F814W ACS images.
Figure~\ref{fig:n2108} shows the F555W ACS image of NGC\,2108, along
with the degraded images and a color image (i.e., a color composite of
F435W, F555W, and F814W ACS images).
% which reveals the locations of blue and red supergiant stars in the
% field.

Additionally, we then perform aperture photometry on the detected
source positions in the non-degraded ACS images using a radius 300
pixels to collect light from all stars that would fall within our
aperture if NGC\,2108 was located at a distance of 5\,Mpc (i.e.,
M83). Objects along the edge and within 1000 pixels of the center of
the cluster in NGC\,2108 are discarded, as would be the case for the
corresponding photometry in M83.  Figure~\ref{fig:circles} shows the
extent of the 300 pixel aperture for several sources in yellow, as
well as the nearest dominant star in red. The magnitude of the
dominant stars (using a 3 pixel aperture and the appropriate aperture
correction) is considered the ``truth'' measurement while the
magnitude in the corresponding 300 pixel aperture is the value that
would be determined at the distance of M83. We calculate the magnitude
and color offsets, i.e., $\Delta I$ and $\Delta (V\!-\!I)$ mag of the
dominant stars, by taking the difference between the 3 pixel and 300
pixel aperture magnitudes.

%This procedure avoids the difficulty of re-calibrating the photometry
%in the degraded images. )
One complication is that our default sky subtraction method would
remove the vast majority of faint and moderate brightness stars in the
background annulus in the NGC\,2108 image, but these stars would not
be detected individually at the distance of M83 and hence would be
included in the sky measurement.  We therefore adjusted the
``clipping'' parameters in the sky subtraction algorithm to remove
only the bright stars in NGC\,2108, thereby mimicking the case for M83
to the degree possible.

In Figure~\ref{fig:m83_reg3} we show the original CMD for Region~3 in
M83 (solid points), compared with a version where the magnitudes and
colors have been `perturbed' by an amount  $\Delta I$ and $\Delta
(V\!-\!I)$ determined by matching to the closest values of \vi\, from
the NGC\,2108 data experiment (open points).  As expected, corrections
for the degraded spatial resolution in M83 tend to make the corrected
magnitudes slightly fainter and redder, as can be seen in
Figure~\ref{fig:m83_reg3}. The mean difference in $V$ is
$-0.24\pm0.30$\,mag with a range from $+0.20$ to $-0.65$\,mag. In $I$,
the mean difference is $-0.40\pm0.39$\,mag with a range from $+0.07$
to $-1.18$\,mag. We note that the largest difference is seen for
object 10, as shown in Figure~\ref{fig:circles}, where several
relatively bright red stars in the 300\,pixel aperture cause the
largest correction (i.e., $-1.18$\,mag in $I$ and redward by 0.63\,mag
in \vi). Object 40, dominated by a single very bright star, shows the
opposite extreme, with a change of $-0.02$\,mag in $V$ and
$-0.01$\,mag in $I$. We also note that object 31 does not make it into
our sample since the two bright stars are far enough apart that it is
clear from the degraded image in Figure~\ref{fig:n2108} that this is
not a single star. This object gets removed in the {\tt DoPHOT}
photometry as described in \S\,2, and hence is not used in the
experiment described above.

We then rerun our age-dating software on the corrected CMD in Figure
\ref{fig:m83_reg3}.  We find a mean age of 23.8\,Myr compared to the
original value of 21.9\,Myr.  Hence, while a specific ``star'' may be
affected by a sizeable amount, the overall effects of degraded spatial
resolution are relatively minor for our study.

To summarize, we have performed a numerical experiment using
observations of an intermediate-age star cluster in the LMC (NGC 2108)
for our ``truth'' image.  Using a factor of 100 in spatial
degradation, roughly appropriate for our measurements in M83, we find
that the typical corrections to our photometry are on the order of a
few tenths of a magnitude, although much larger values are possible in
specific cases. We conclude that while spatial resolution can result
in measurable differences in the luminosities and colors of bright
stars at the distance of M83, it does not strongly affect the age
estimates for the 50~regions studied in this paper.

%            EQUATION
%--------------------------------------------------
%Equation example:
%\begin{equation}\label{av_def}
%  A_V = (m_V - m_{V,0})
%\end{equation}

%            REFERENCE
%--------------------------------------------------
%\clearpage

\clearpage

%            FIGURES
%--------------------------------------------------
\begingroup
% Figure 1
\begin{figure}
%\centerline{\psfig{file=m83_reg50_colbox_092911.eps,width=1.0\textwidth}}
\begin{center}
\includegraphics[width=\textwidth]{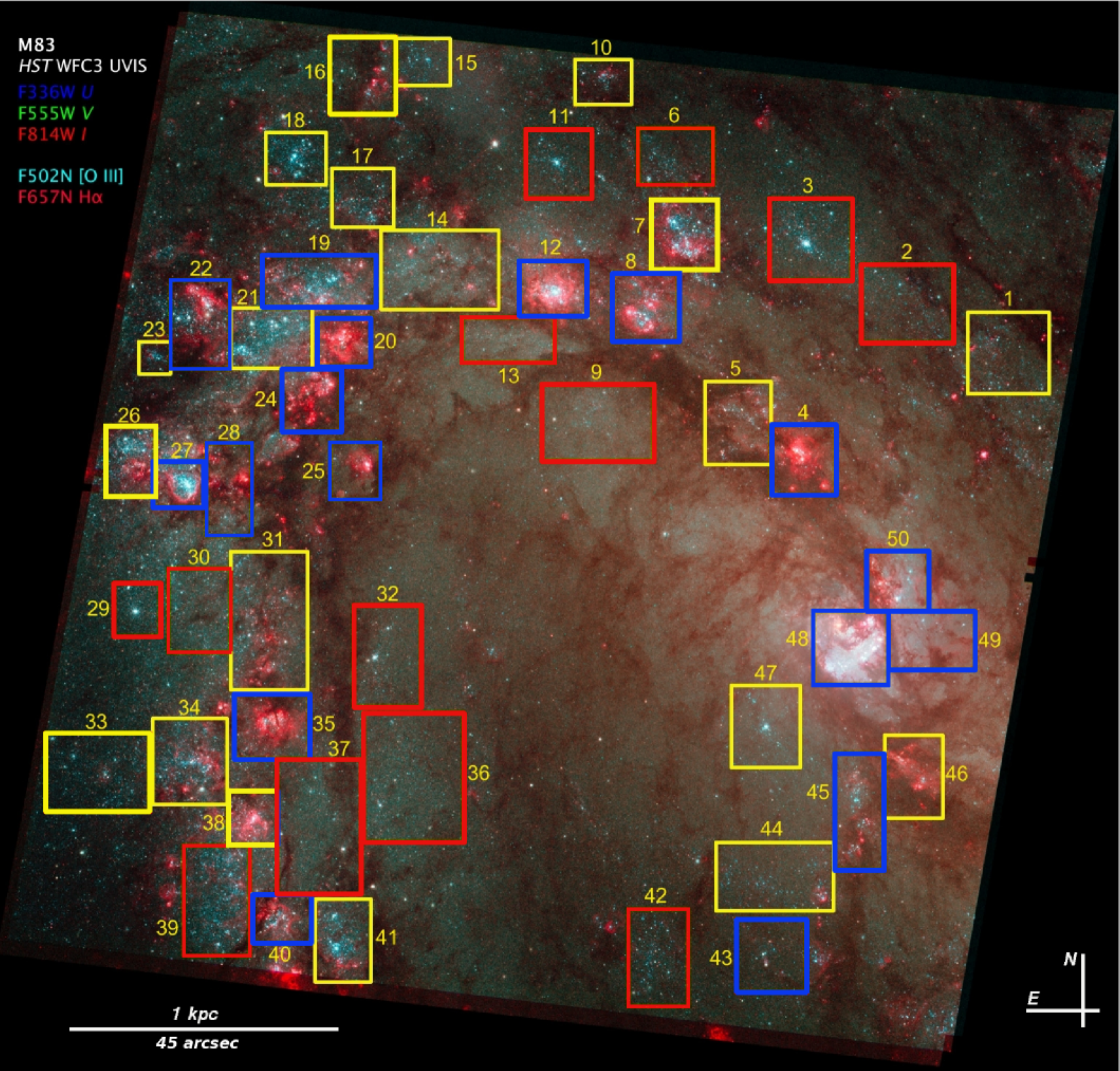}
\caption{A color composite of the M83 WFC3 images [Image Credit: Zolt
  Levay (STScI)]. The F336W image is shown in blue, the F502N ([O
  III]) image in cyan, the F555W image in green, and the combined
  F814W and F657N ($H\alpha$) image in red.  The 50~selected regions
  are outlined in boxes of three different colors based on the values
  of $Age_{CMD}$$_{auto}$ determined in this paper (see \S3.1 for
  details). The regions with ages in the range 1--10\,Myr are outlined
  in blue, ages of 10--20\,Myr in yellow, and ages greater than
  20\,Myr in red.}
\label{fig:m83_reg50}
\end{center}
\end{figure}

\newpage

% Figure 2
\begin{figure}
\begin{center}
\includegraphics[width=\textwidth]{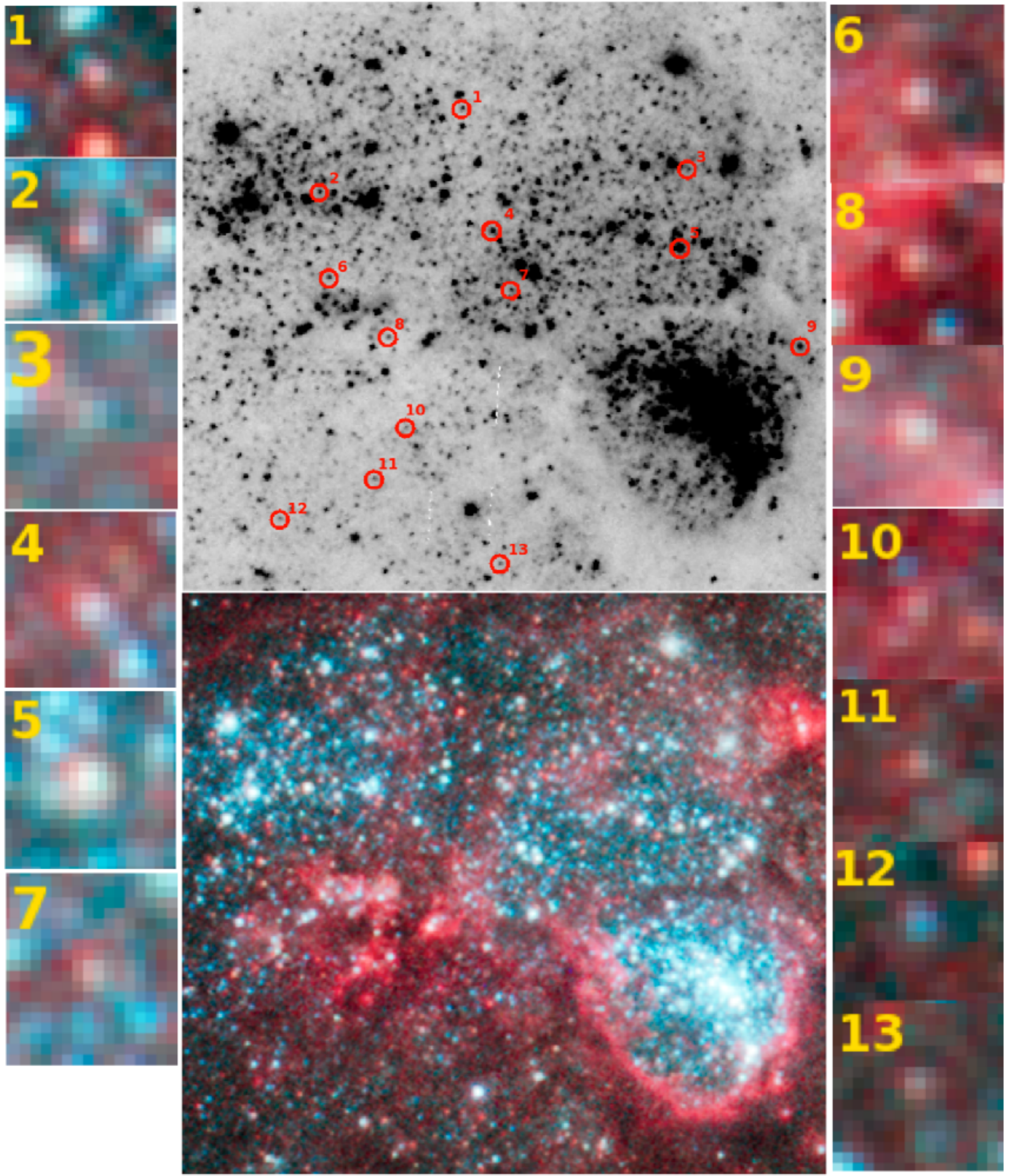}
%\centerline{\psfig{file=pseudo_fig.eps,width=0.95\textwidth}}
\caption{Image cut-outs (15$\farcs$2$\times$13$\farcs$9,
  340\,pc$\times$310\,pc) of Regions \#~26 and \#~27 in the F555W
  ($top$) and the color-composite images ($bottom$). Stars circled in
  red have $(V-I)>$0.8 mag. The ``postage stamp'' images on the left
  are``pseudo'' matched candidates, while the images on the right
  appear to be heavily reddened (see \S2.2 for details).}
\label{fig:pseudo}
\end{center}
\end{figure}

% physical scale of images are calculated based on pixel scale of
% 0.885 pc/pixel at D=4.61 Mpc.
% 385x385 pixel = 

% Figure 3
\begin{figure}
\centering
\subfigure{\includegraphics[width=\textwidth]{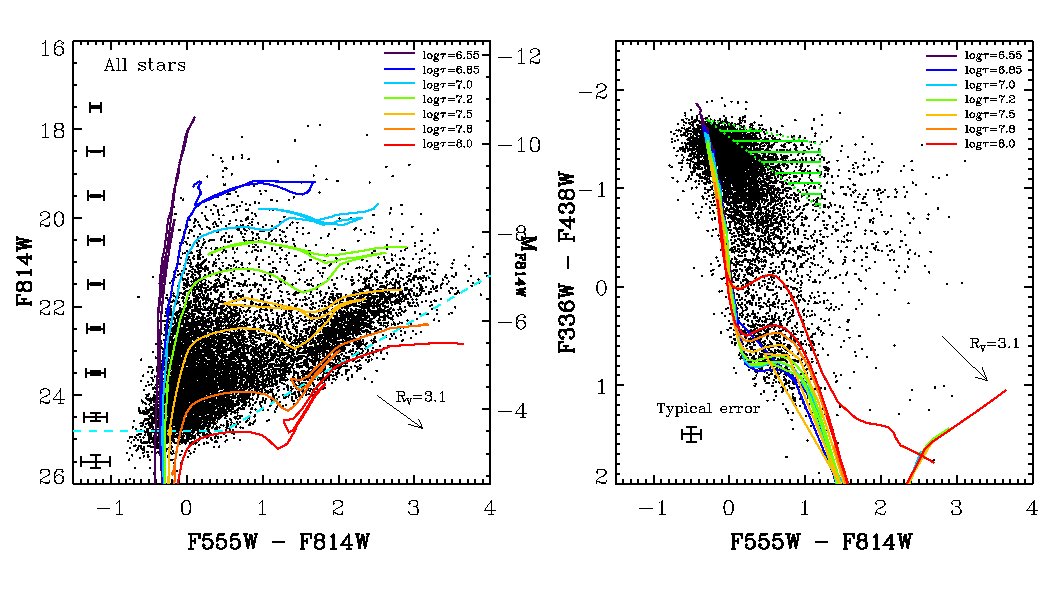}}
\subfigure{\includegraphics[width=0.45\textwidth]{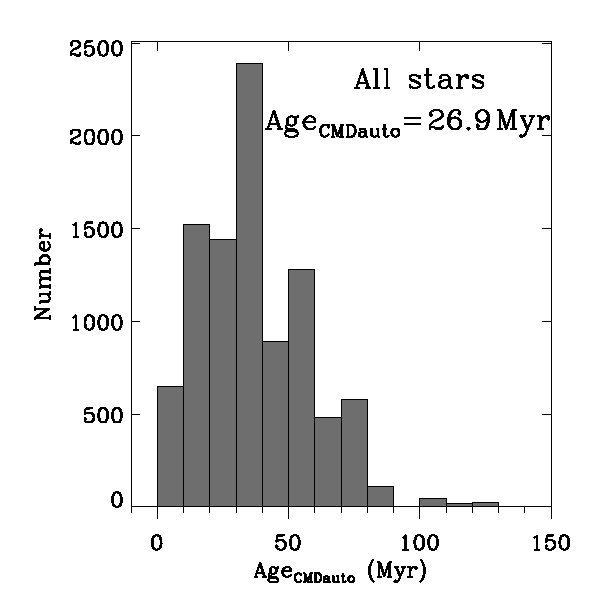}}
\subfigure{\includegraphics[width=0.45\textwidth]{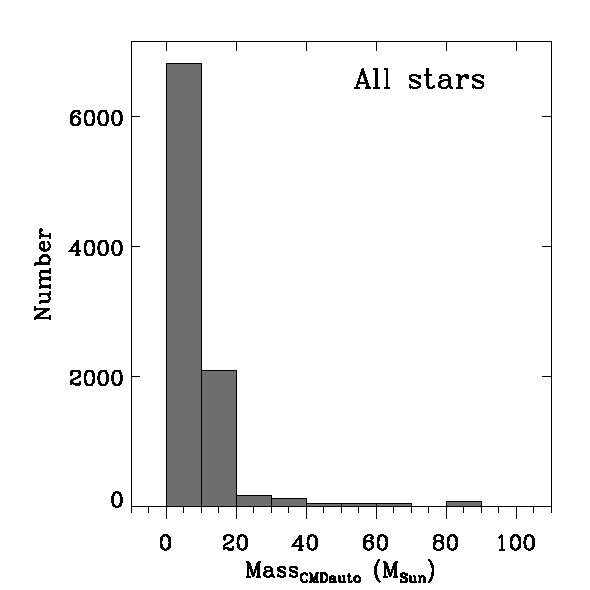}}
\caption{{\it Top:} The color-magnitude diagram (CMD) and color-color
  diagram of all stars in our M83 image corrected for Galactic
  foreground extinction but not for internal extinction. Padova
  isochrones of log$\tau$ (age/yr) = 6.55, 6.85, 7.0, 7.2, 7.5, 7.8,
  and 8.0 for a metallicity of Z=0.03 (1.5 Z$_{\odot}$) are overlaid
  in both panels. The dashed line in cyan represents the 50\%
  photometric completeness level. The arrow in each panel indicates
  the direction of the reddening vector with $R_V=3.1$. Magnitudes are
  on the Vega scale. The distance modulus $(m-M)_{0} = 28.32$ mag is
  used to calculate the absolute magnitude $M_{F814W}$. {\it Bottom:}
  Histograms of the distribution of ages and masses, as determined
  from the CMD (see \S 3.1 for details). The age quoted in the bottom
  left panel is the luminosity-weighted mean age.}
\label{fig:allstar_no}

\end{figure}

% Figure 4
\begin{figure}
\centering
\subfigure{\includegraphics[width=\textwidth]{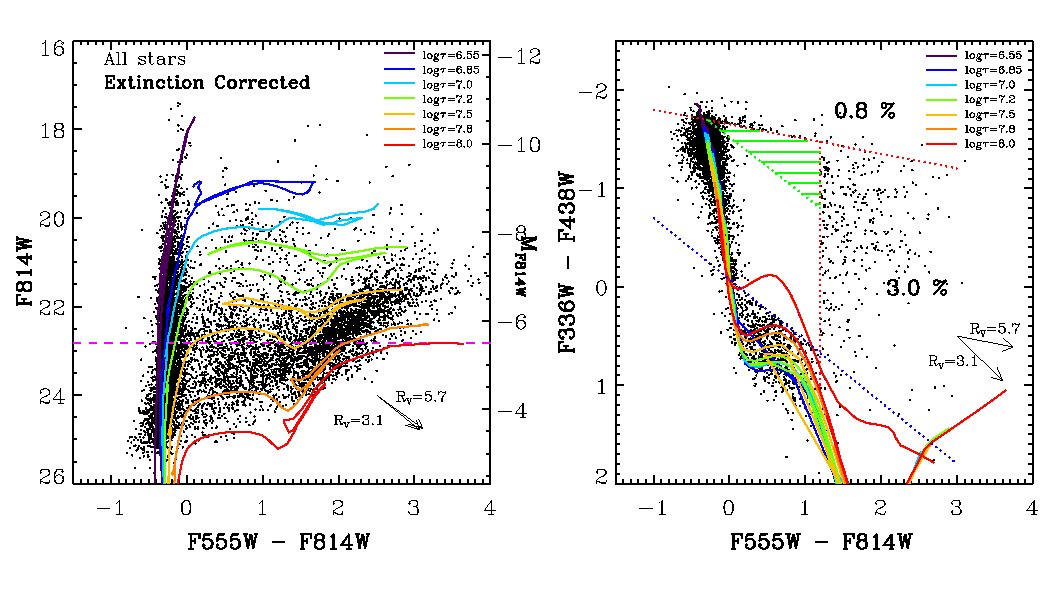}}
\subfigure{\includegraphics[width=0.45\textwidth]{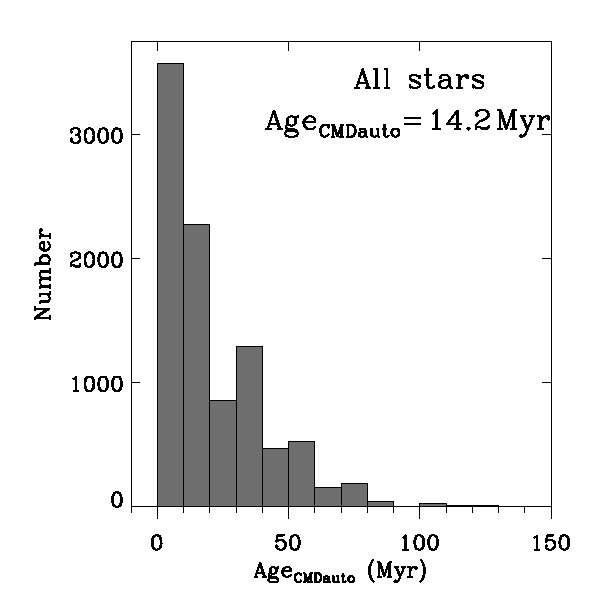}}
\subfigure{\includegraphics[width=0.45\textwidth]{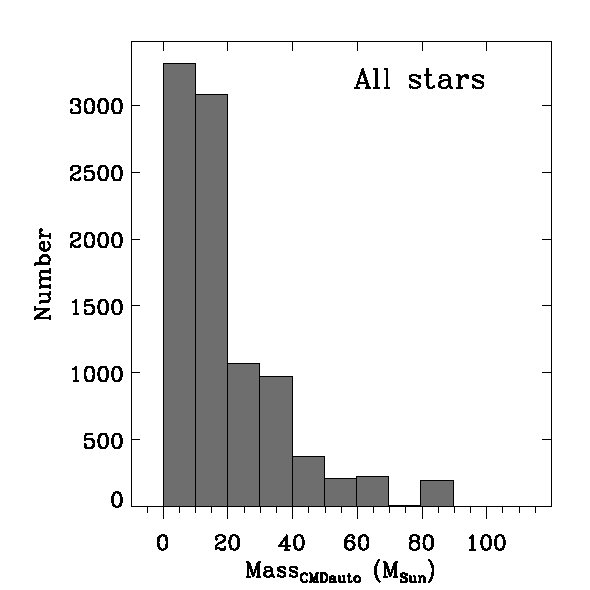}}
\caption{{\it Top:} The CMD and color-color diagram of all stars
  corrected for the internal extinction determined for each individual
  star. The blue and two red dotted lines indicate boundaries for the
  locations of stars in the color-color diagram uncorrected for
  internal extinction (see \S 2.4 for discussion). Two arrows in the
  color-color diagram show reddening vectors: one with the standard
  reddening vector ($R_{V}=3.1$) and the other with the flatter
  reddening vector ($R_{V}=5.7$) which appears to be more appropriate
  for some of the data in the green triangle (see \S 2.4 for
  discussion). The dashed line in magenta shows the $M_{I}$$<$$-5.5$
  magnitude cutoff used for the $Age_{CMDauto}$ estimate. {\it
    Bottom:} Same as described in Fig. \ref{fig:allstar_no}.}
\label{fig:allstar_cor}
\end{figure}
\newpage

% Figure 5
\begin{figure}
\centering
\subfigure{\includegraphics[width=0.2\textwidth]{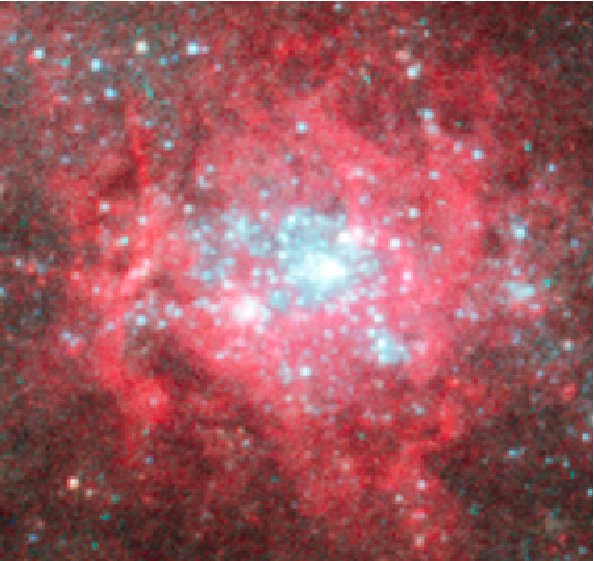}}
\subfigure{\includegraphics[width=0.6\textwidth]{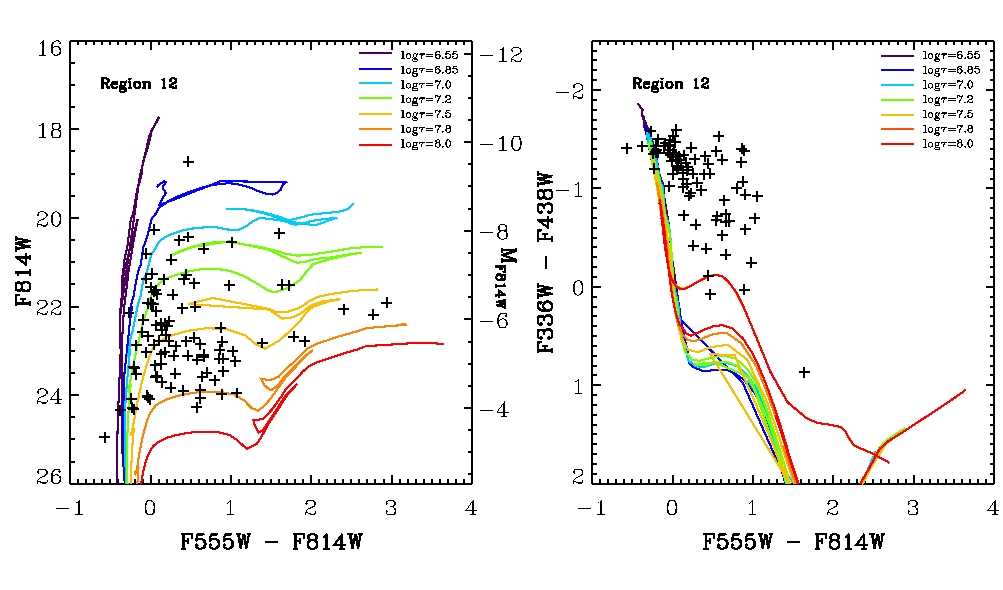}}
\subfigure{\includegraphics[width=0.2\textwidth]{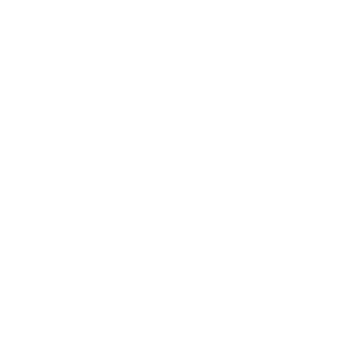}}
\subfigure{\includegraphics[width=0.6\textwidth]{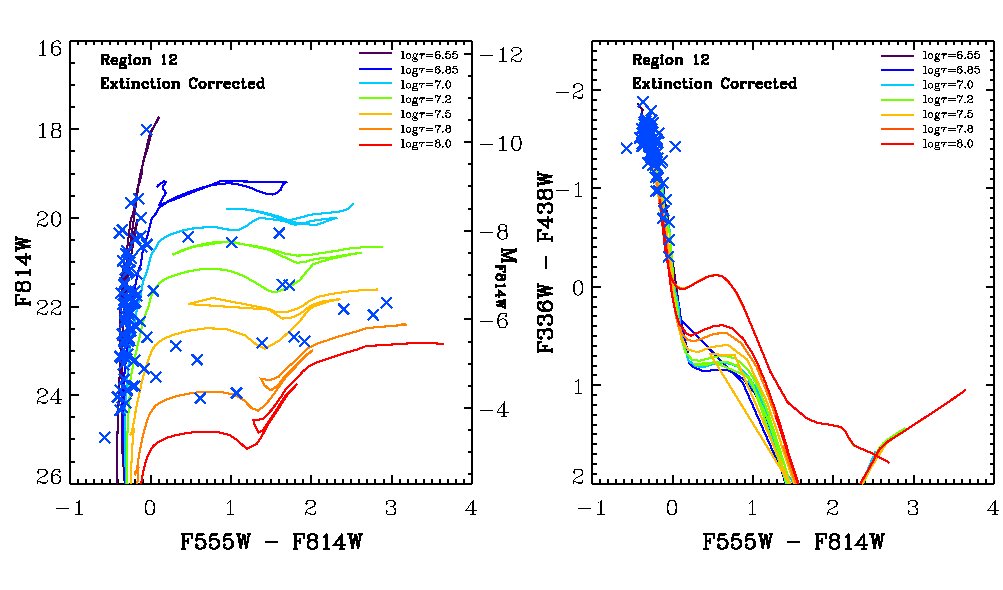}}

\centering
\subfigure{\includegraphics[width=0.2\textwidth]{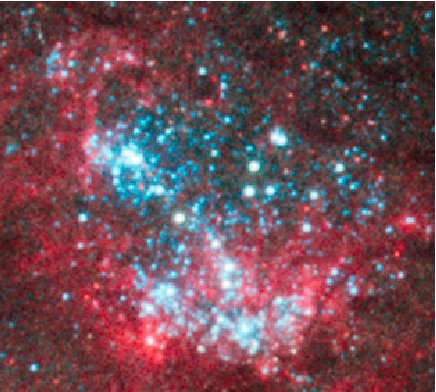}}
\subfigure{\includegraphics[width=0.6\textwidth]{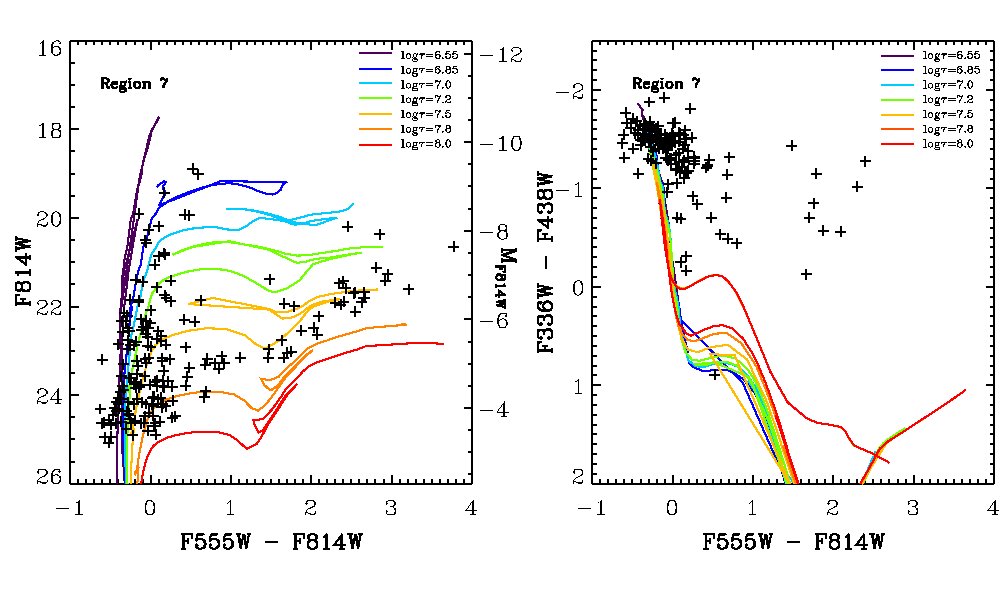}}
\subfigure{\includegraphics[width=0.2\textwidth]{blank.jpg}}
\subfigure{\includegraphics[width=0.6\textwidth]{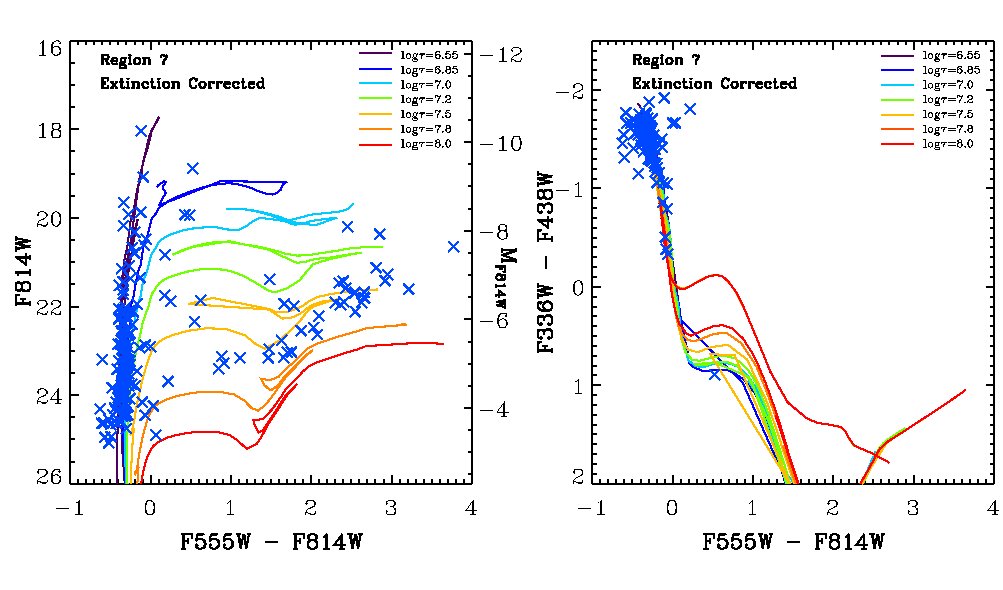}}

      \caption{The CMDs, color-color diagrams and the image cut-outs
        of Regions \#~12 and \#~7. Data points before and after the
        individual extinction correction are plotted as black crosses
        (upper: uncorrected) and blue Xs (lower: corrected). The
        Padova isochrones in these diagrams are the same as the
        ones plotted in Figure \ref{fig:allstar_no}. See \S 2.4
        for details.}
\label{fig:reg12_7}
\end{figure}

% Figure 6 : Figure 4-cont'd
\begin{figure}
\centering
\subfigure{\includegraphics[width=0.2\textwidth]{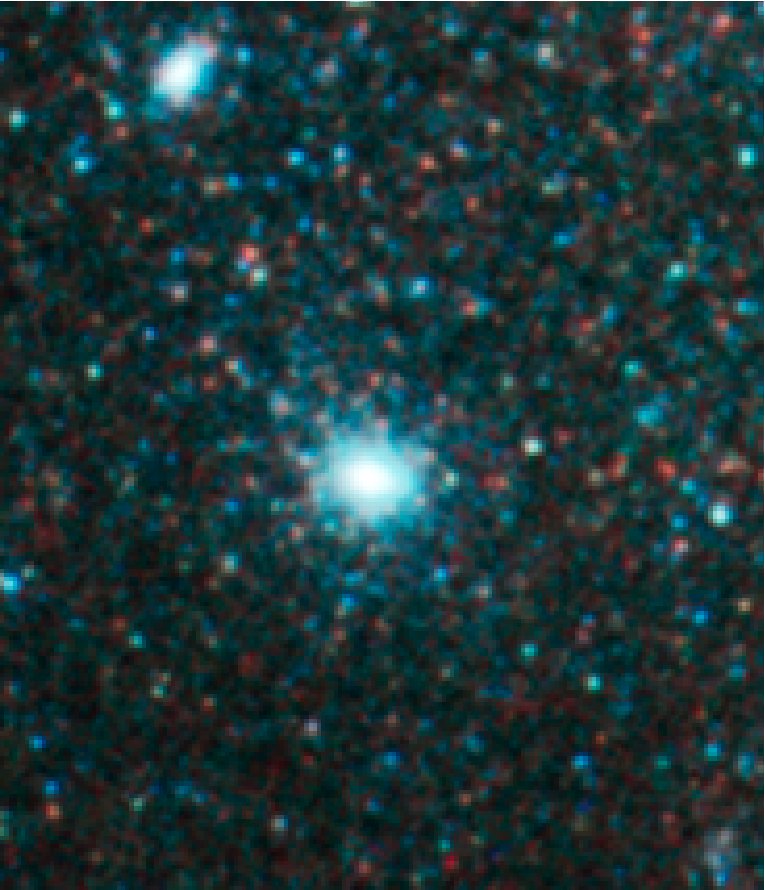}}
\subfigure{\includegraphics[width=0.6\textwidth]{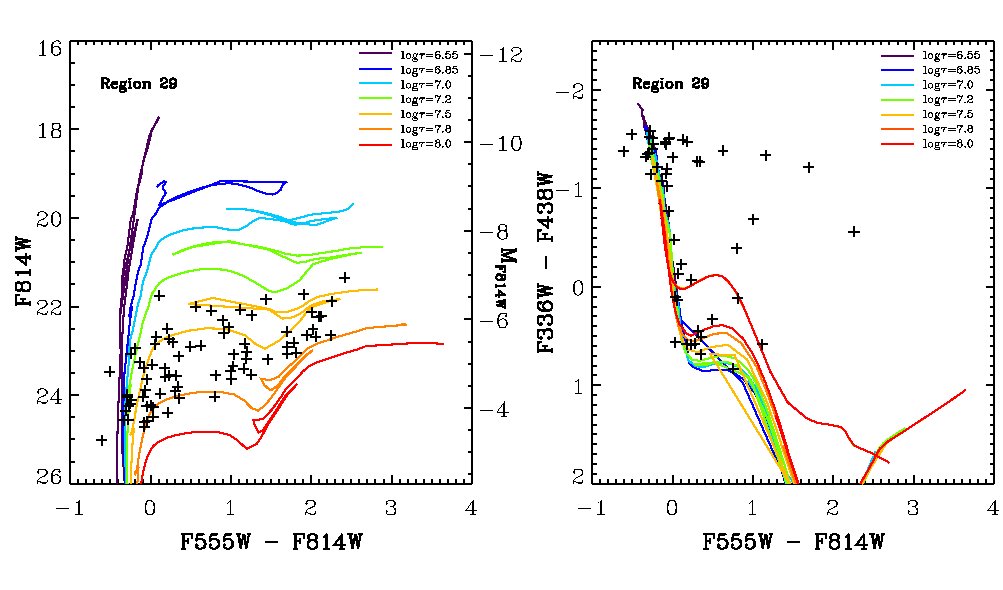}}
\subfigure{\includegraphics[width=0.2\textwidth]{blank.jpg}}
\subfigure{\includegraphics[width=0.6\textwidth]{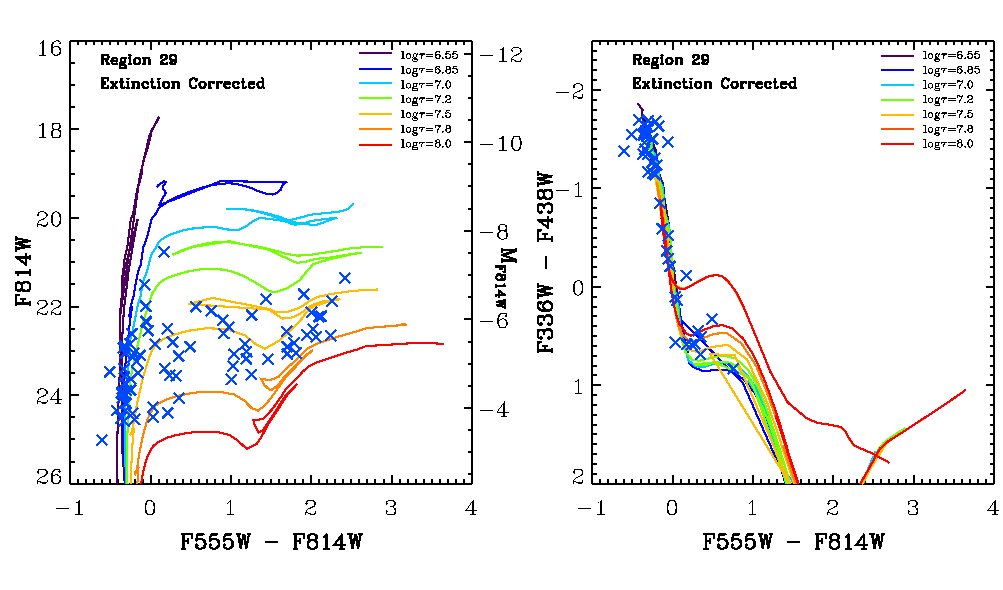}}

\centering
\subfigure{\includegraphics[width=0.2\textwidth]{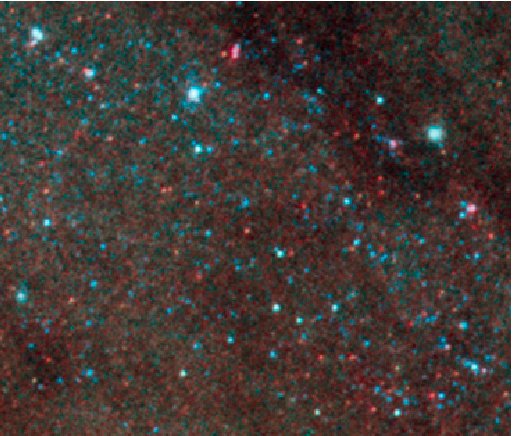}}
\subfigure{\includegraphics[width=0.6\textwidth]{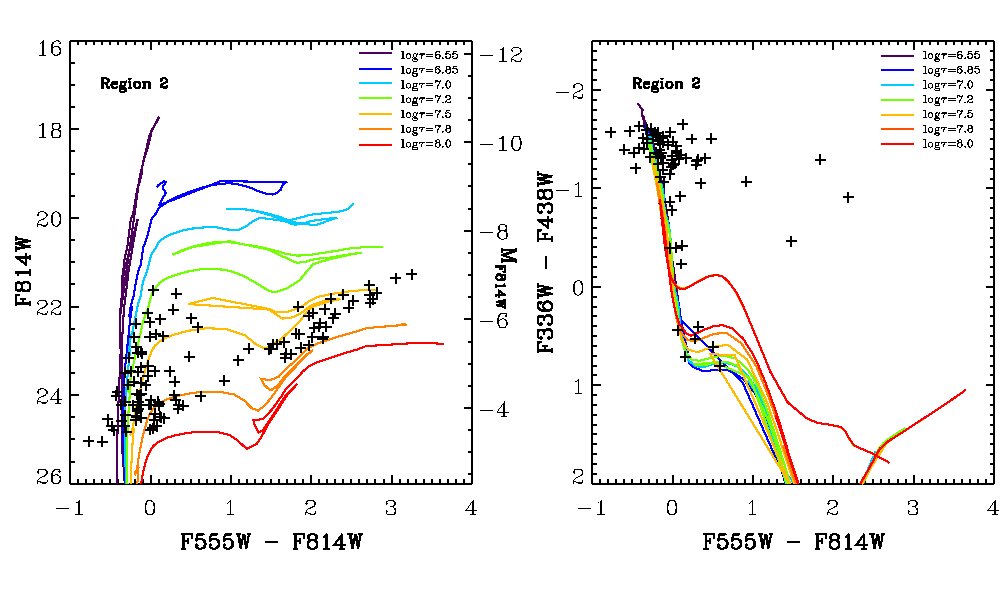}}
\subfigure{\includegraphics[width=0.2\textwidth]{blank.jpg}}
\subfigure{\includegraphics[width=0.6\textwidth]{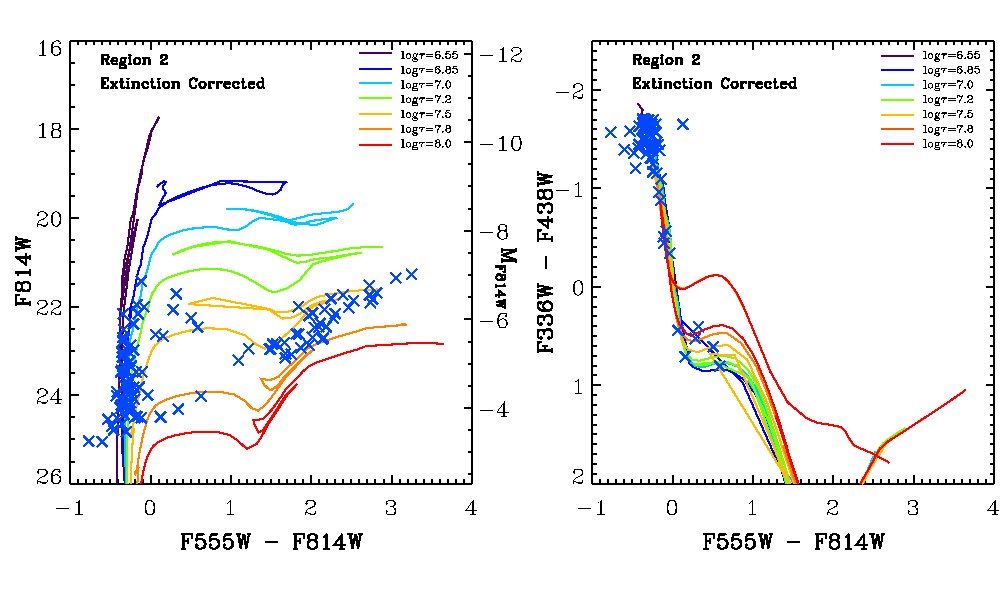}}

\caption{Same as Figure \ref{fig:reg12_7} for Regions \#~29 and \#~2.}
\label{fig:reg29_2}
\end{figure}

% Figure 8
\begin{figure}
\centering
\subfigure{\includegraphics[width=0.45\textwidth,height=0.33\textheight]{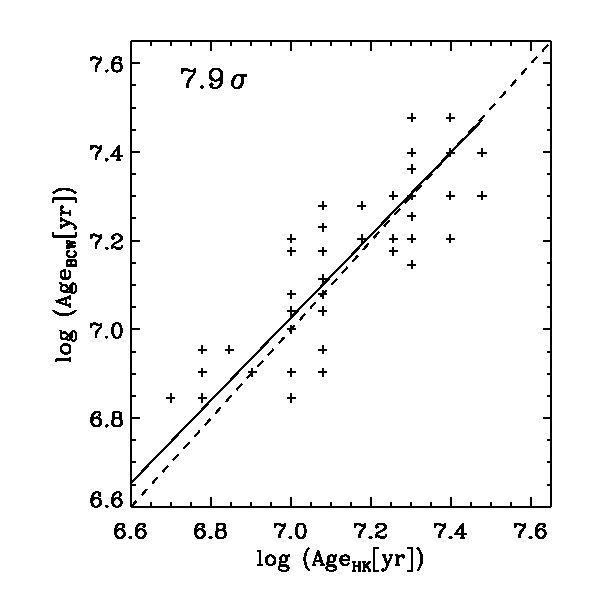}}
\subfigure{\includegraphics[width=0.45\textwidth,height=0.33\textheight]{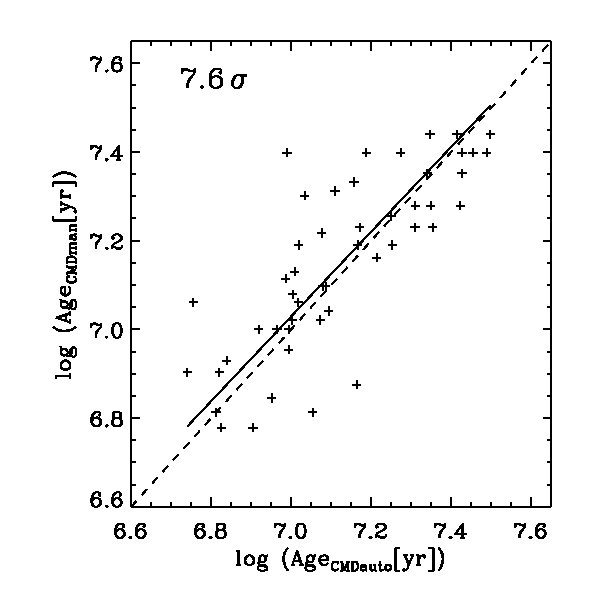}}
\caption{Correlation plots between the ages of the stellar populations
  in the 50~regions from CMD isochrone fitting. $(Left)$: Correlation
  between the manual ages determined by HK and BCW. $(Right)$:
  Correlation between the average of manual ages and automatic ages
  (see \S 3.1 for details). The best linear fits are shown in black
%for the 47 regions
with the significance of the correlation in unit of $\sigma$ (i.e.,
slope/uncertainty) in the top left of each panel.
% Regions 48, 49, and 50 (marked as diamonds) are excluded in the
%  linear fitting due to the photometric incompleteness in the regions
%  near the nucleus.
  The dashed line is the unity line. }
\label{fig:cmdages}
\end{figure}

% Figure 7
\begin{figure}
\centering
\subfigure{\includegraphics[width=0.45\textwidth,height=0.25\textheight]{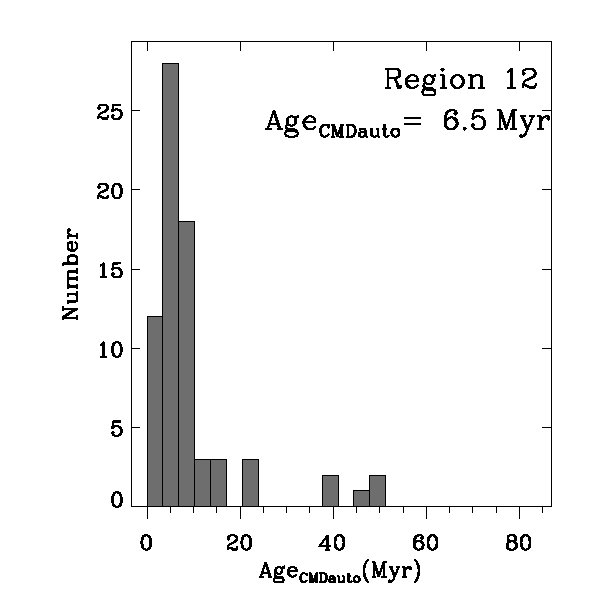}}
\subfigure{\includegraphics[width=0.45\textwidth,height=0.25\textheight]{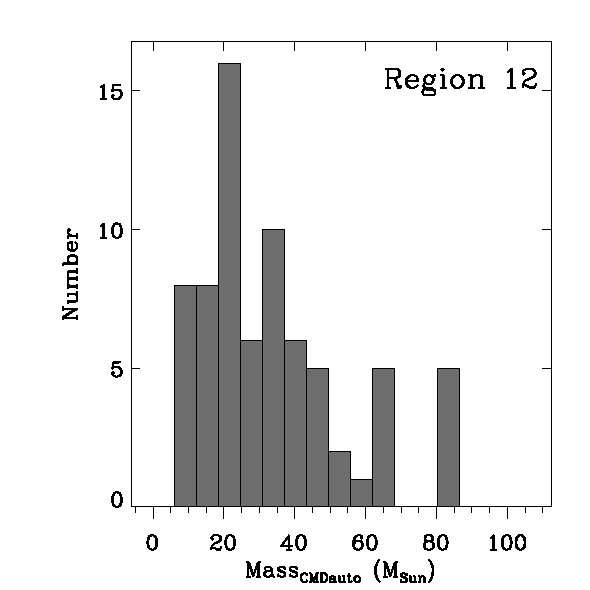}}
\subfigure{\includegraphics[width=0.45\textwidth,height=0.25\textheight]{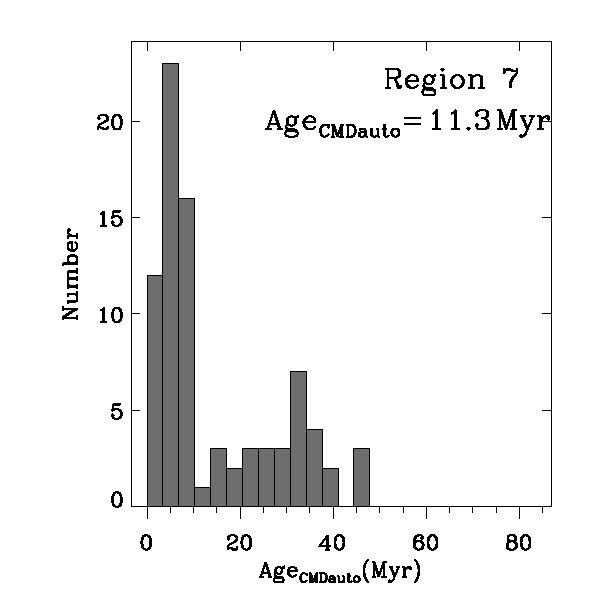}}
\subfigure{\includegraphics[width=0.45\textwidth,height=0.25\textheight]{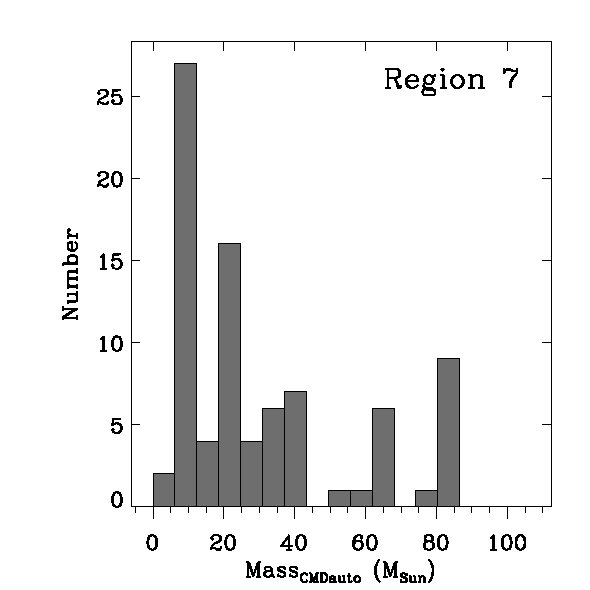}}
\subfigure{\includegraphics[width=0.45\textwidth,height=0.25\textheight]{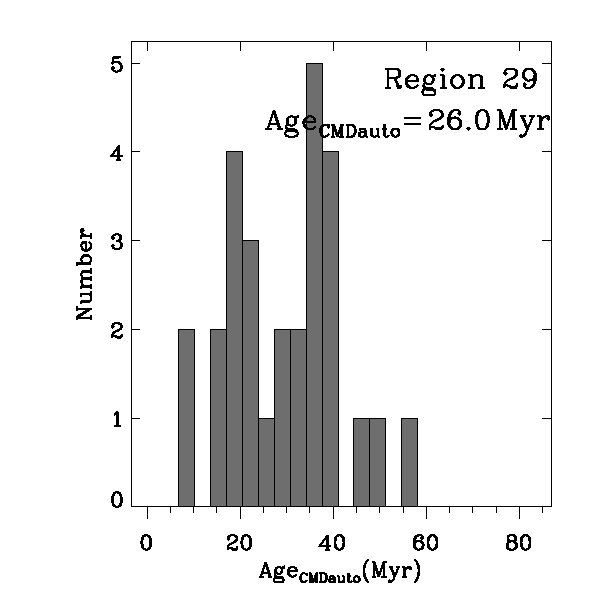}}
\subfigure{\includegraphics[width=0.45\textwidth,height=0.25\textheight]{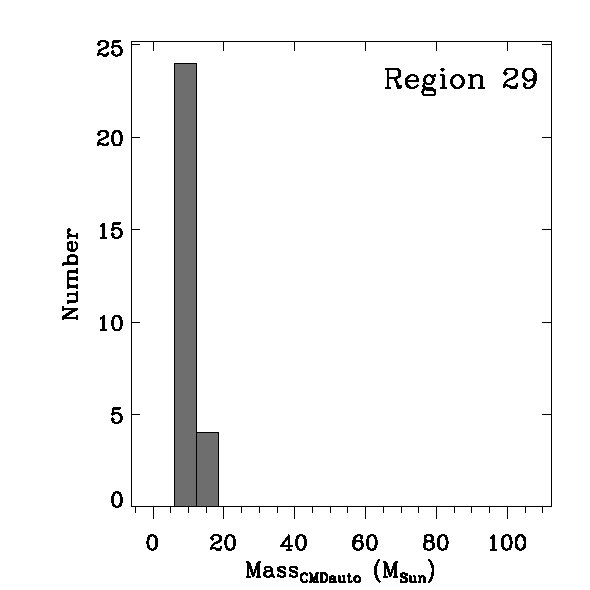}}
\subfigure{\includegraphics[width=0.45\textwidth,height=0.25\textheight]{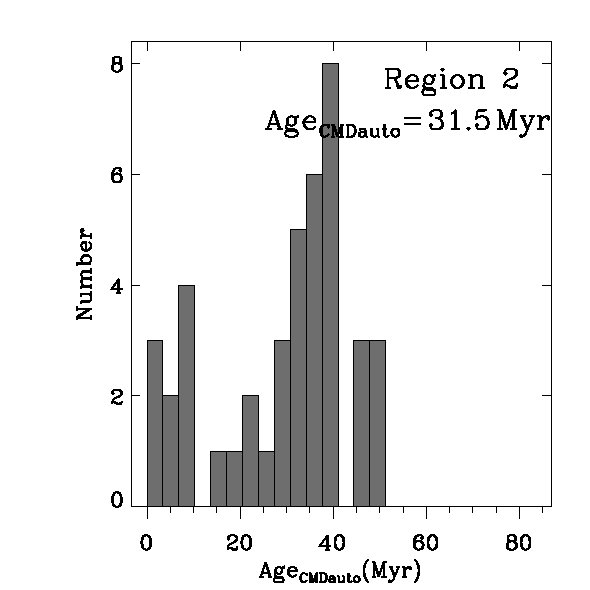}}
\subfigure{\includegraphics[width=0.45\textwidth,height=0.25\textheight]{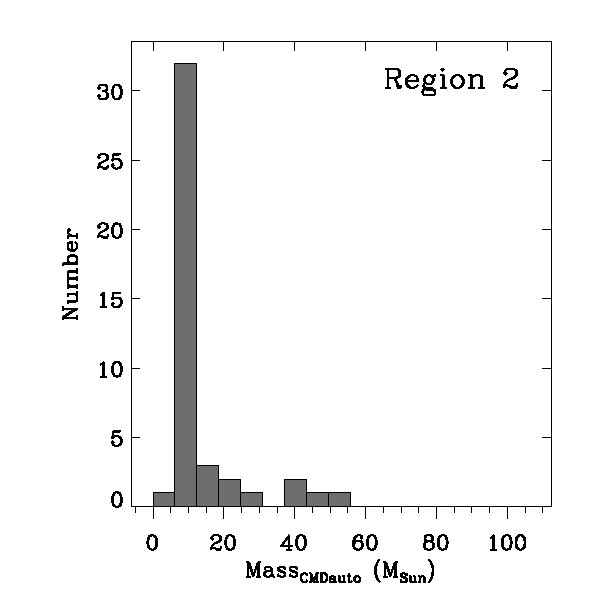}}

\caption{Histograms of the distribution of ages and masses of the
  individual stars in Regions \#~12, \#~7, \#~29, and \#~2, as determined by the
  automated isochrone fitting technique (see \S 3.1). The age
  indicated at the top right in each age histogram is the
  luminosity-weighted mean age of all stars with $M_{I}$$<$$-5.5$ in each region.}
\label{fig:age_mass}
\end{figure}

% Figure 9
\begin{figure}
\centering
\includegraphics[width=0.5\textwidth]{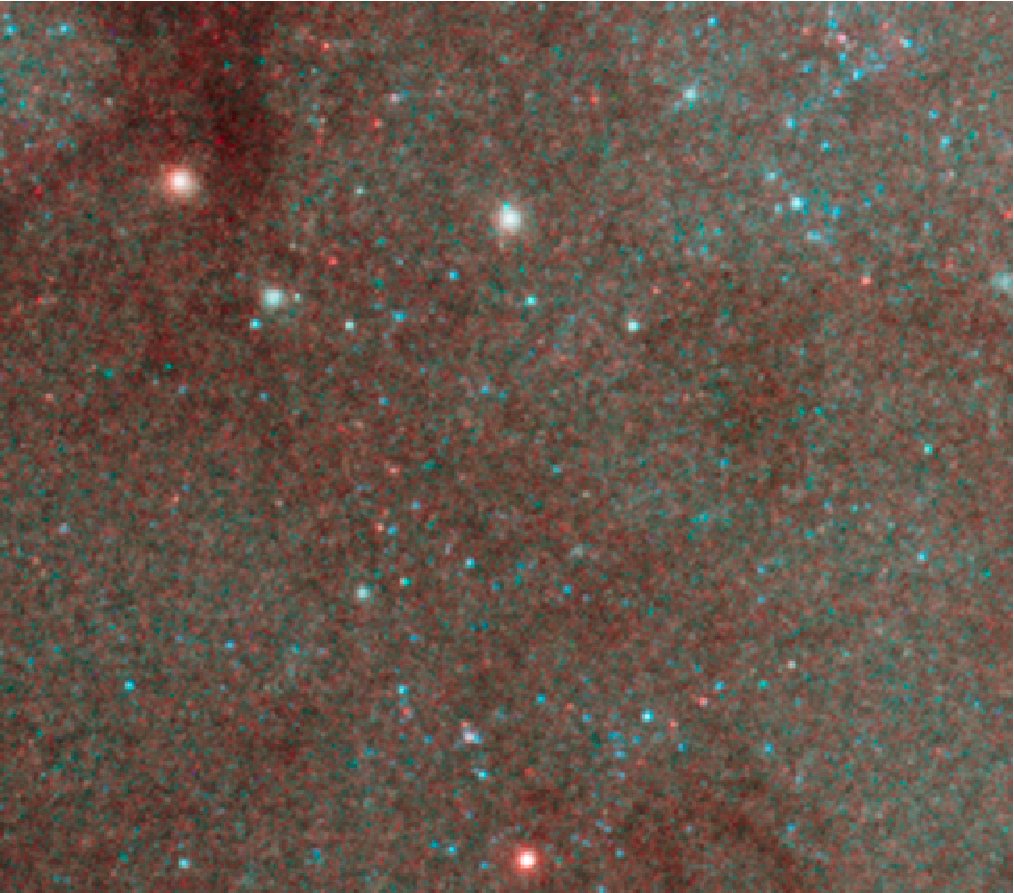}
\caption{Image cutout (18$\farcs$3$\times$16$\farcs$9,
  410\,pc$\times$380\,pc) of the area near Region 9 as an example of a
  ``field'' region with old red giant stars, and a small population of
  isolated young blue stars. See \S 3.1 for for detail. }
\label{fig:reg9}
\end{figure}

% 460x430 pixel = 410pcx380

% Figure 10
\begin{figure}
\centering
\includegraphics[height=0.7\textheight]{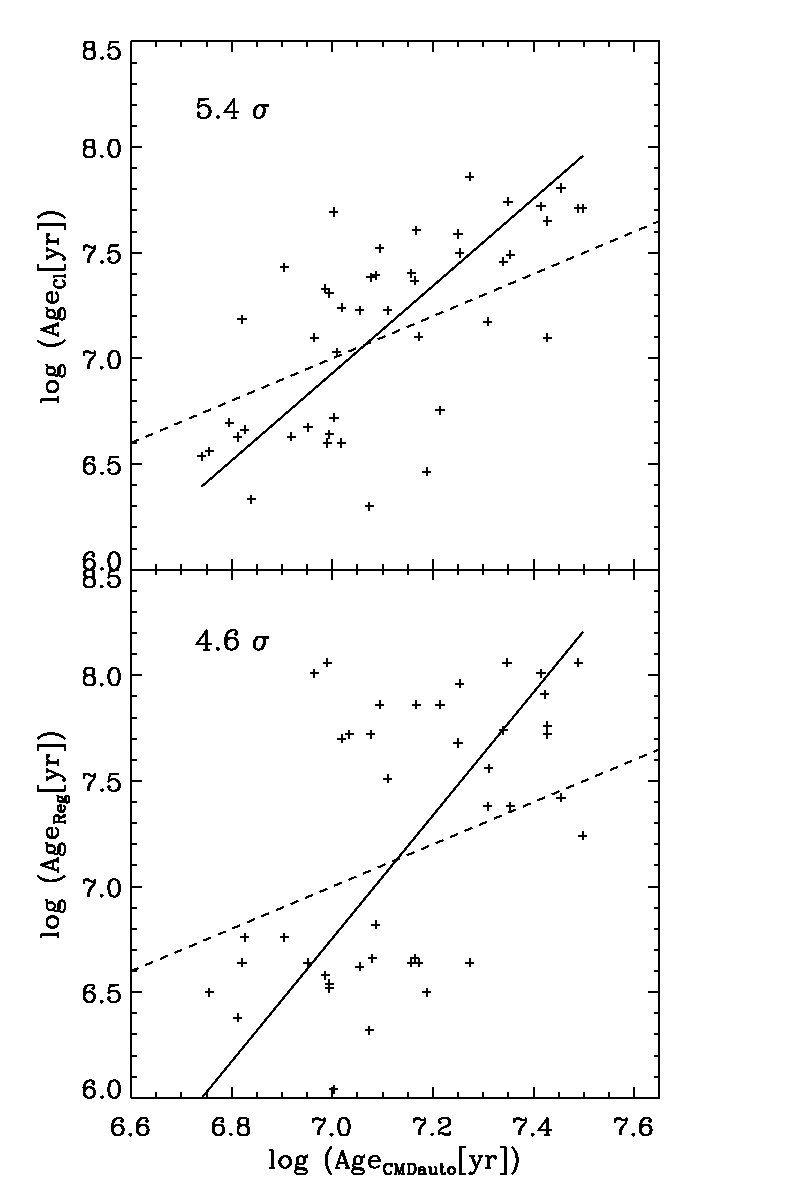}
\caption{($Top$): Correlations between the ages of stars
  ($Age_{CMD}$$_{auto}$) and clusters ($Age_{Cl}$). ($Bottom$):
  Correlation between the ages of stars determined by
  $Age_{CMD}$$_{auto}$ and the integrated light within the region
  ($Age_{Reg}$). See \S 3.4 for details.
%The stellar ages are luminosity-weighted ages, determined using the
%  automatic method and plotted on x-axis in log(age/yr). The ages of
%  the star clusters ($top$: the mean ages of stars clusters in each
%  region; and $middle$: the luminosity-weighted ages of star clusters
%  in each region) and the regions ($bottom$): the region ages using
%  the integrated photometry of each region), determined from the
%  Spectral Energy Distribution (SED) fitting are plotted on y-axis.
  The best linear fits are shown in black solid line with
  5.4~$\sigma$ $(top)$ and 4.6~$\sigma$ $(bottom)$ correlations.
% Three regions near the nucleus of M83 are excluded to fit the line
% (marked as diamonds).
  The dashed line is the unity line.}
\label{fig:sedages}
\end{figure}

% Figure 11
\begin{figure}
\centering
\includegraphics[height=0.9\textheight]{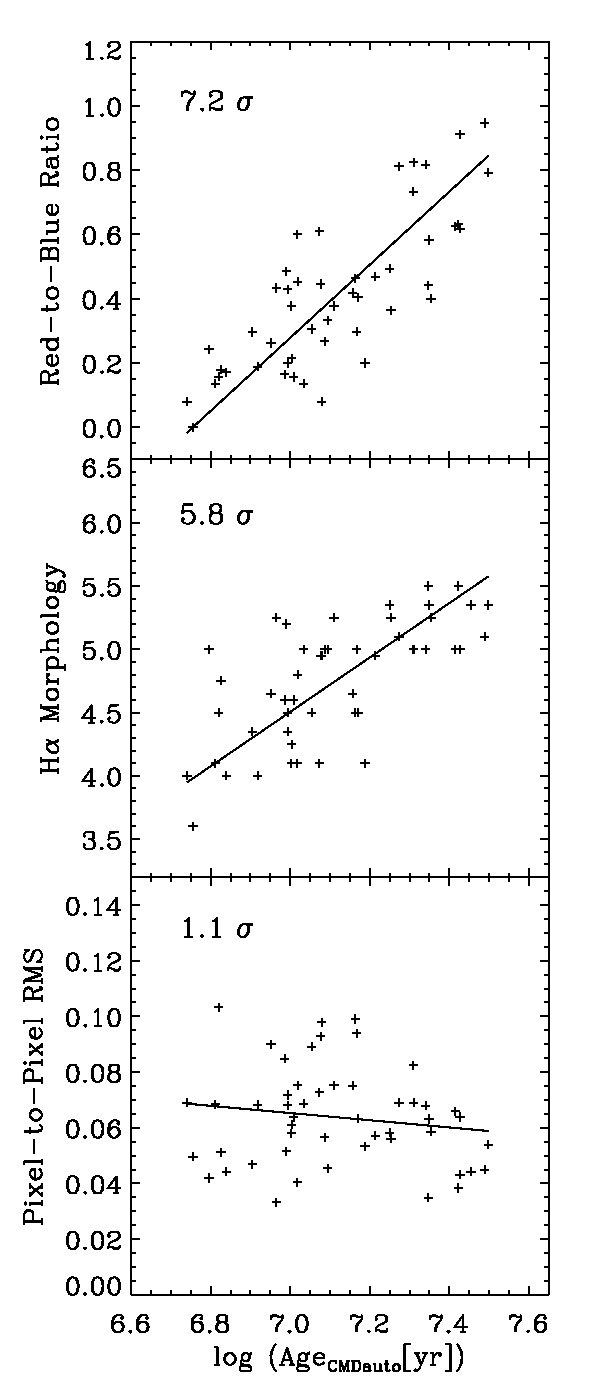}
\caption{Correlations between the stellar ages ($Age_{CMD}$$_{auto}$) and
  the number ratio of red-to-blue stars $(top)$, \ha morphology $(middle)$,
  and pixel-to-pixel fluctuations $(bottom)$ for
  the 50 selected regions. See \S 4 for details.
% The solid black lines are the best linear fits for the 49 regions
%  with $\sim9\sigma$, $\sim6\sigma$, and $\sim1\sigma$
%  correlations. The region \# 48 is excluded in all three plots.
  The best linear fits are indicated in solid black line with
  7.2~$\sigma$ $(top)$, 5.8~$\sigma$ $(middle)$, and 1.1~$\sigma$
  $(bottom)$ correlations.}

\label{fig:rb_ha_rms}
\end{figure}

% Figure 12
\begin{figure}
\centering
\includegraphics[height=0.92\textheight]{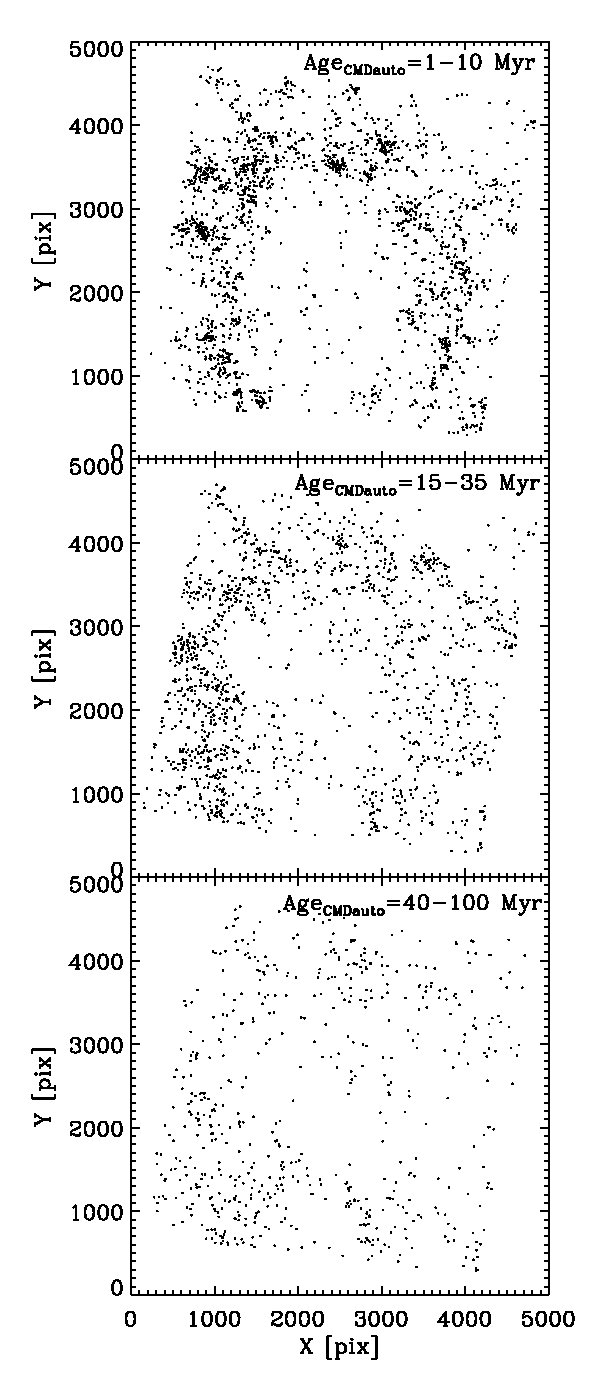}
\caption{Spatial distribution of stars with ages of 1--10, 15--35, and
  40--100 Myr determined in this study. Each panel has the same
  orientation and scale as Figure 1. The top panel is the youngest
  group of stars, clearly showing that the stars in these regions are
  mostly distributed along the active star-forming region (i.e.,
  associated with the strong \ha emission) in the spiral arms. The
  stars in the middle panel tend to be found slightly downstream of
  the spiral arms while the older stars are still farther out in the
  inter-arm regions, as expected. See \S 5 for details.}
\label{fig:pellerin}
\end{figure}

% Figure 13
\begin{figure}
\centering
\includegraphics[width=1.0\textwidth]{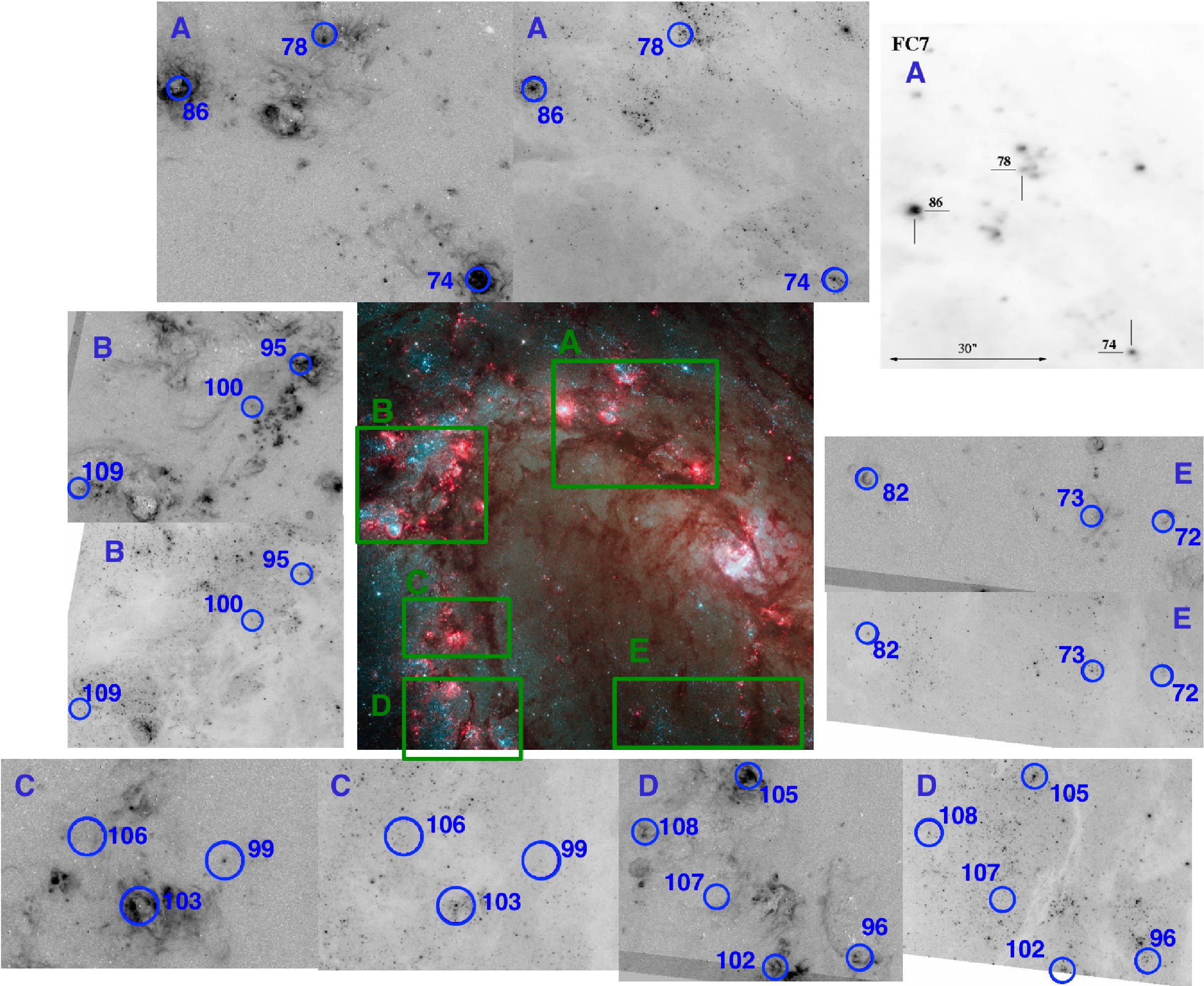}
\caption{Distribution of Wolf-Rayet sources in our \hst\/ WFC3 M83
  images. The color image at the center is the $UBVIH\alpha$ composite
  and the image cut-outs are the \ha ($left$ or $top$) and F814W
  ($right$ or $bottom$) images for regions A--E.
The top right panel is an original finding chart
    \citep[][$\lambda$4684 narrow-band FORS2 image]{hadfield05} for
    the region A.
Numbers in blue are the IDs of Wolf-Rayet sources identified by
\citet{hadfield05}.}
\label{fig:wr1}
\end{figure}

% Figure 14
\begin{figure}
\centering
\includegraphics[width=1.0\textwidth]{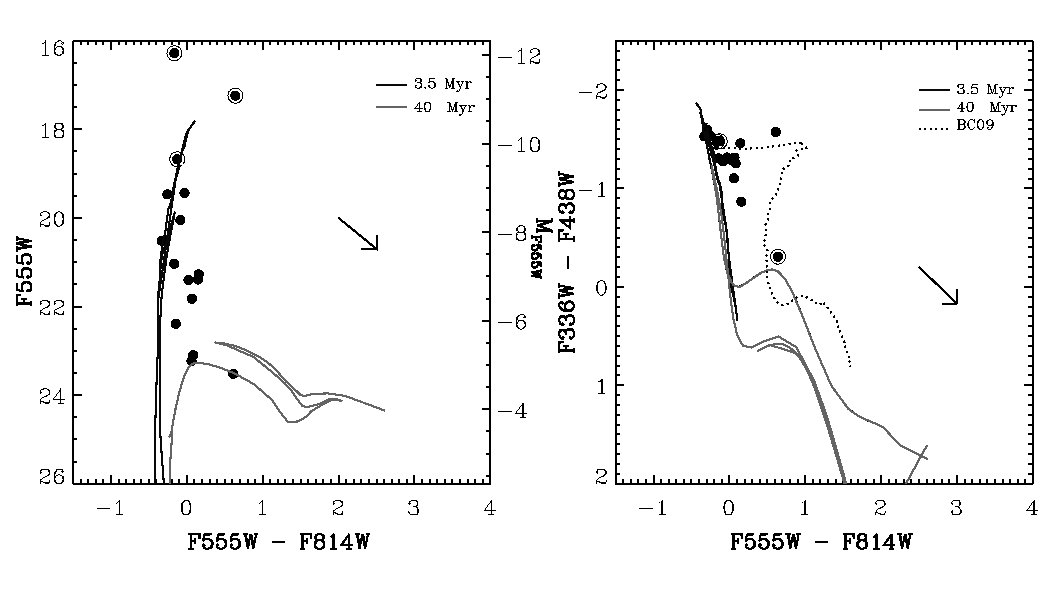}
\caption{The CMD and color-color diagram of Wolf-Rayet sources in M83
  marked in Figure \ref{fig:wr1}. Three sources (\#\,78, 86, 105)
  circled in both diagrams are identified as clusters in our M83 WFC3
  images.  A twice solar metallicity BC09 cluster model (Bruzual \&
Charlot 2009, private communication; see also Bruzual \& Charlot 2003)
is shown in the right panel (dots), and Padova stellar models
\citep{marigo08} are shown in both panels. No correction has been made
for internal reddening, which appears to be minimal for all but one
object. The arrow indicates the Galactic reddening vector.}
\label{fig:wr2}
\end{figure}

% Figure 15
\begin{figure}
\centering
\includegraphics[width=1.0\textwidth]{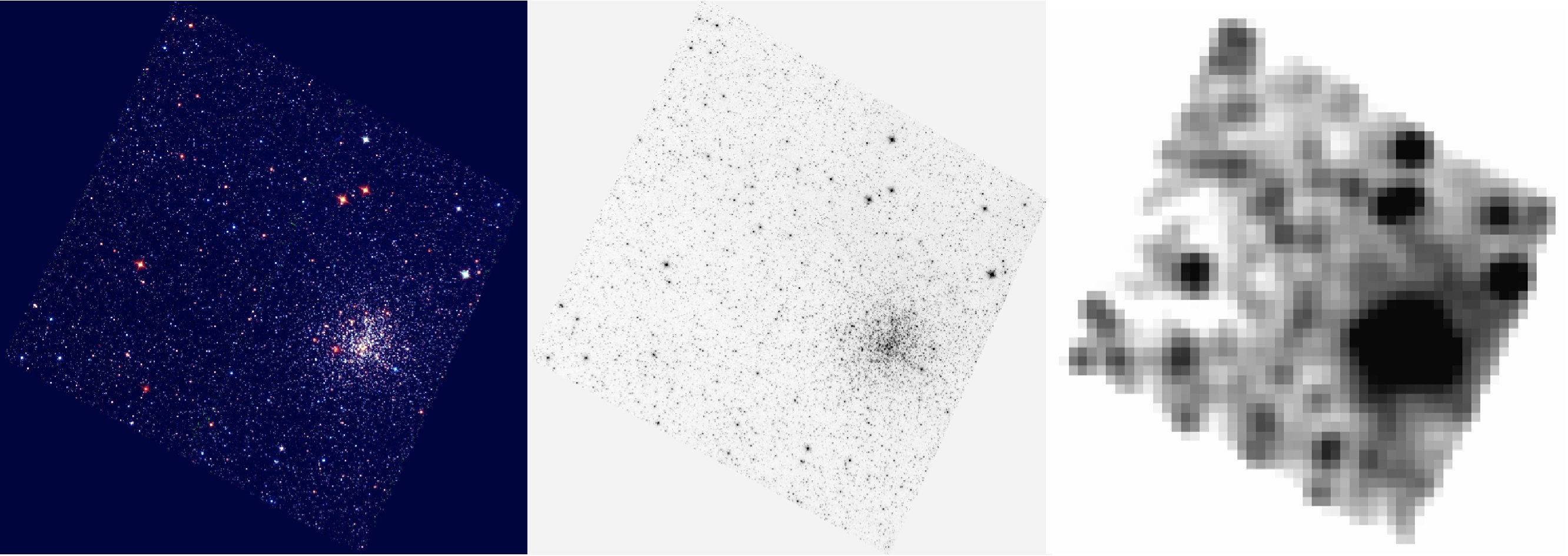}
\caption{$HST$/ACS images (290$\arcsec\times$290$\arcsec$,
  68\,pc$\times$68\,pc) of the compact star cluster NGC\,2108 in the
  LMC: a color composite of the F435W, F555W, and F814W ACS images
  ($left$), the F555W ACS image ($middle$), and the degraded F555W ACS
  image ($right$) by a factor of 100 (from a distance of $\sim$50\,kpc
  to a distance of $\sim$5\,Mpc). See Appendix B for details.} 
\label{fig:n2108}
\end{figure}

%ngc2108 physical scale: D= 48.977 kpc, ACS pixel scale = 0.05 arcsec
% (m-M)_0 = 18.45 
% 5800x5800 pixel = 290x290 arcsec = 68pc x 68pc 

% Figure 16
\begin{figure}
\centering
\subfigure{\includegraphics[width=0.8\textwidth]{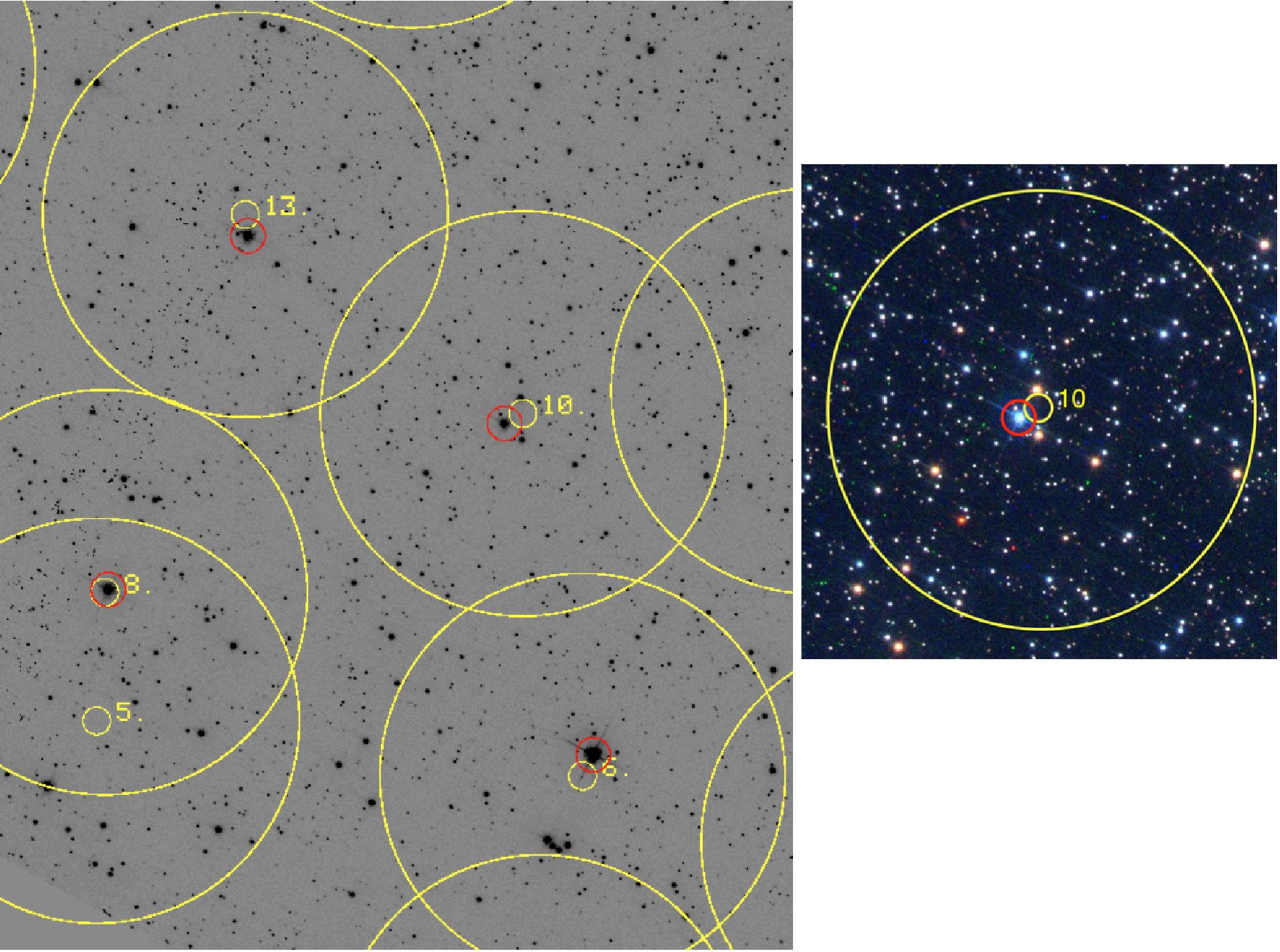}}
\subfigure{\includegraphics[width=0.7\textwidth]{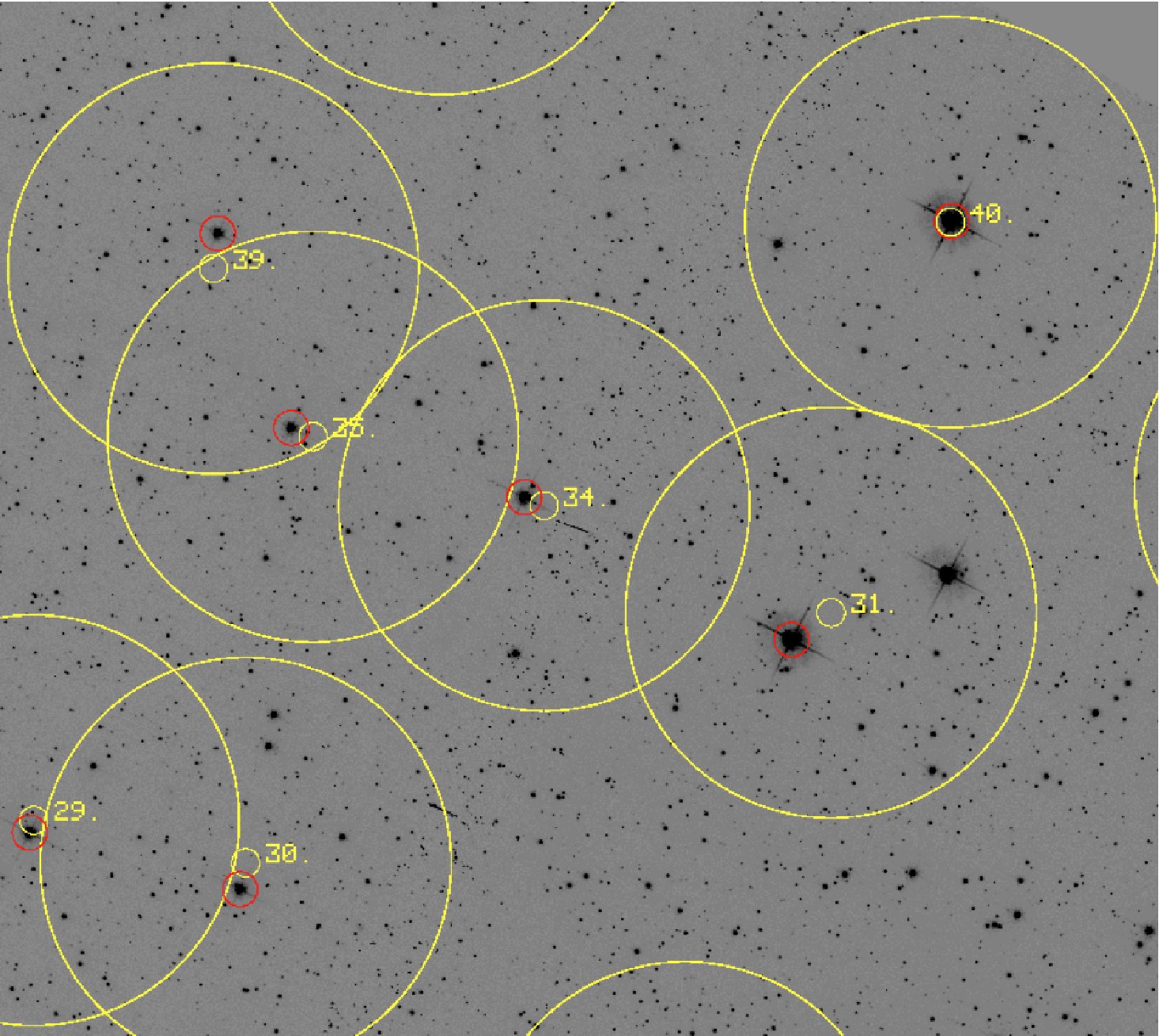}}
\caption{Image cutouts of \hst/ACS F555W image of NGC\,2108. The large
  circles in yellow show the 300-pixel aperture (corresponding to a
  $\sim$3\,pixel aperture at the distance of M83) centered at the
  location determined from the degraded image in Figure 15. The
  location of the dominant star (used to determine the ``truth''
  value) is shown in red.  See Appendix B for details.}
\label{fig:circles}
\end{figure}

% Figure 17
\begin{figure}
\centering
\includegraphics[width=0.8\textwidth]{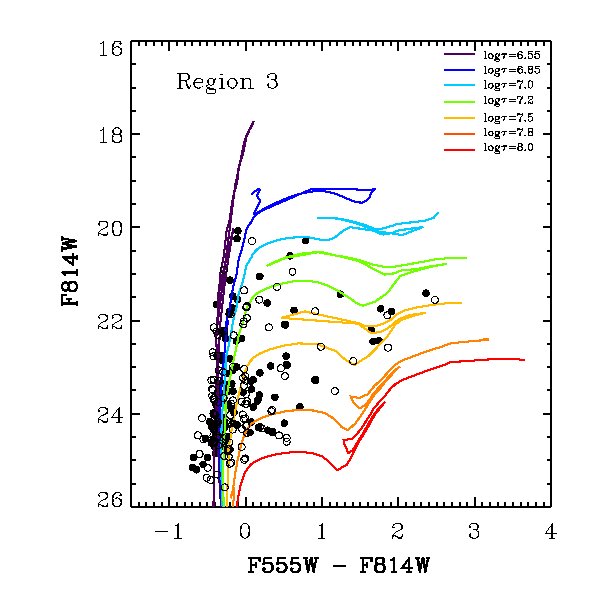}
\caption{The original CMD of Region\,3 in M83 (solid points) with the
  effect of simulation representing a factor of 100 degration in
  spatial resolution (open points). The $Age_{CMD}$$_{auto}$ of this
  region changes from 21.9\,Myr (determined from the solid points) to
  23.8\,Myr (from the open points). The difference in the age
  estimates is within $\sim1\,\sigma$ error listed in Table 1. See
  Appendix B for details.}
\label{fig:m83_reg3}
\end{figure}

\endgroup

\clearpage
%            TABLE 
%--------------------------------------------------
%Table 1.  Properties of 50~regions
\addtocounter{footnote}{100}
\begingroup
\tiny
\centering
\begin{sidewaystable}
%\begin{landscape}
\begin{longtable}{lllcccccccccccc}
\multicolumn{15}{c}{Table 1. Properties of the 50~regions} \\
%\caption{Table 1. Properties of 50~regions}\\
\hline \hline
\multicolumn{1}{c}{Region} &
\multicolumn{1}{c}{$\!\!\!$R.A.(J200)} &
\multicolumn{1}{c}{$\!\!\!$Dec.(J2000)} &
\multicolumn{1}{c}{$\!\!\!$Region Size} &
\multicolumn{1}{c}{$\!\!\!$$Age_{CMD}$$_{man}$} &
\multicolumn{1}{c}{$\!\!\!$$Age_{CMD}$$_{auto}$} &
\multicolumn{1}{c}{$\!\!\!$$\Delta$$Age^{a}$} &
\multicolumn{1}{c}{$\!\!\!$$Age_{Reg}^{b}$} &
%\multicolumn{1}{c}{$\!\!\!$$Age_{Cl}$\footnotemark{a}} &
\multicolumn{1}{c}{$\!\!\!$$Age_{Cl}^{c}$} &
\multicolumn{1}{c}{$\!\!\!$Pixel-to-Pixel} &
\multicolumn{1}{c}{$\!\!\!$Red-to-blue} &
\multicolumn{1}{c}{$\!\!\!$$H\alpha$} &
\multicolumn{1}{c}{$\!\!\!$$A_{V}^{d}$} &
\multicolumn{1}{c}{$\!\!\!$N$_{Star}$}&
\multicolumn{1}{c}{$\!\!\!$N$_{Cl}$} \\
\multicolumn{1}{c}{} &
\multicolumn{1}{c}{$\!\!\!$(hh:mm:sec)} &
\multicolumn{1}{c}{$\!\!\!$(\arcdeg:':'')} &
\multicolumn{1}{c}{$\!\!\!$(pc$\times$pc)} &
\multicolumn{1}{c}{$\!\!\!$(Myr)} &
\multicolumn{1}{c}{$\!\!\!$(Myr)} &
\multicolumn{1}{c}{$\!\!\!$(Myr)} &
\multicolumn{1}{c}{$\!\!\!$(Myr)} &
\multicolumn{1}{c}{$\!\!\!$(Myr)} &
\multicolumn{1}{c}{RMS} &
\multicolumn{1}{c}{Ratio} &
\multicolumn{1}{c}{Morphology} &
\multicolumn{1}{c}{(Mag)} &
\multicolumn{1}{c}{} &
\multicolumn{1}{c}{}
\\ \hline
\endfirsthead

\endhead

\multicolumn{15}{l}{$^a$$\Delta$$Age_{CMD}$$_{auto}$ is the mean value
  of the uncertainties in our age determination. See \S 3.1 for
  details. } \\

\multicolumn{15}{l}{$^b$$Age_{Reg}$ is the region age determined by
  measuring the colors of the entire regions, and fitting SED.  See \S
  3.4 for details.}\\

\multicolumn{15}{l}{$^c$The luminosity-weighted mean age of clusters
  in each region. Age of each cluster is determined by SED
  fitting. See \S 3.3 for details.}\\

\multicolumn{15}{l}{$^d$ Average value of the extinctions ($A_{V}$)
  measured from the resolved stars in each region. } \\

\multicolumn{15}{l}{$^e$Region \#~48 is the nucleus of M83. It is not
  included in the analysis since the bright background results in a
  completeness limit that is more than two magnitudes }\\

\multicolumn{15}{l}{brighter than other regions.} \\

\endlastfoot

1 & 13:36:59.78 & -29:52:30.19 & 310$\times$310 & 18.0 & 17.8 & 1.8 & 47.9 & 38.8 & 0.0582 & 0.493 & 5.35 & 0.288 & 206 & 3 \\
2 & 13:37:01.11 & -29:52:21.60 & 350$\times$290 & 27.5 & 31.5 &  2.2 & 17.4 & 51.3 & 0.0539 & 0.792 & 5.35 & 0.214 & 231 & 2 \\
3 & 13:37:02.40 & -29:52:10.72 & 310$\times$310 & 22.5 & 21.9 &  1.7 & 28.7 & 28.7 & 0.0679 & 0.817  & 5.00 & 0.161 & 311 & 5 \\
4 & 13:37:02.48 & -29:52:48.35 & 250$\times$260 & 8.5 & 6.9 &  0.5 & --- & 2.2 & 0.0442 & 0.171 & 4.00 & 0.858 & 67 & 4 \\
5 & 13:37:03.35 & -29:52:42.20 & 240$\times$290 & 12.5 & 12.2 &  1.0 & 6.6 & 24.8 & 0.0566 & 0.267 & 5.00 & 0.677 & 174 & 2 \\
6 & 13:37:04.18 & -29:51:56.32 & 280$\times$210 & 17.0 & 22.6 &  1.1 & 24.0 & 30.9 & 0.0585 & 0.400 & 5.25 & 0.289 & 132 & 1 \\
7 & 13:37:04.07 & -29:52:09.72 & 250$\times$250 & 6.5 & 11.3 &  1.2 & 4.2 & 16.9 & 0.0891 & 0.306 & 4.50 & 0.305 & 261 & 9 \\
8 & 13:37:04.58 & -29:52:22.24 & 250$\times$260 & 9.0 & 9.9 &  0.5 & 3.5 & 20.3 & 0.0718 & 0.200 & 4.35 & 0.369 & 198 & 7 \\
9 & 13:37:05.22 & -29:52:41.88 & 420$\times$280 & 27.5 & 22.3 &  2.9 & 114.8 & --- & 0.0349 & 0.442 & 5.50 & 0.542 & 121 & 0  \\
10 & 13:37:05.15 & -29:51:43.48 & 210$\times$170 & 13.5 & 10.2 &  2.6 & --- & 10.7 & 0.0638 & 0.156 & 4.60 & 0.726 & 74 & 3 \\
11 & 13:37:05.72 & -29:51:57.44 & 240$\times$250 & 25.0 & 26.7 & 1.0 & 52.5 & 12.6 & 0.0639 & 0.911 & 5.00 & 0.186 & 174 & 2 \\
12 & 13:37:05.80 & -29:52:18.88 & 250$\times$200 & 6.5 & 6.5 &  0.3& 2.4 & 4.2 & 0.0684 & 0.136 & 4.10 & 0.758 & 133 & 4 \\
13 & 13:37:06.42 & -29:52:27.48 & 340$\times$180 & 19.0 & 26.5 &  1.8 & 81.2 & --- & 0.0383 & 0.632 & 5.50 & 0.378 & 79 & 0 \\
14 & 13:37:07.31 & -29:52:15.96 & 440$\times$310 & 15.5 & 17.9 &  0.9 & 91.2 & 31.6 & 0.0562 & 0.364 & 5.25 & 0.475 & 323 & 5 \\
15 & 13:37:07.53 & -29:51:40.04 & 200$\times$180 & 20.0 & 10.8 &  0.5 & 52.5 & --- & 0.0686 & 0.135 & 5.00 & 0.322 & 93 & 0 \\
16 & 13:37:08.32 & -29:51:42.44 & 240$\times$300 & 25.0 & 15.4 &  0.6 & 3.2 & 2.9 & 0.0534 & 0.200 & 4.10 & 0.408 & 108 & 2 \\
17 & 13:37:08.31 & -29:52:03.56 & 230$\times$230 & 20.5 & 12.9 &  2.0 & 32.4 & 16.9 & 0.0752 & 0.377 & 5.25 & 0.439 & 180 & 4 \\
18 & 13:37:09.21 & -29:51:56.48 & 230$\times$190 & 16.5 & 11.9 &  0.3 & 52.5 & 24.2 & 0.0928 & 0.446 & 4.95 & 0.204 & 176 & 6 \\
19 & 13:37:08.91 & -29:52:17.48 & 430$\times$180 & 7.0 & 8.9 &  0.5 & 4.4 & 4.7 & 0.0900 & 0.262 & 4.65 & 0.392 & 288 & 5 \\
20 & 13:37:08.62 & -29:52:27.07 & 180$\times$180 & 10.0 & 8.3 &  0.4 & 0.8 & 4.3 & 0.0682 & 0.188 & 4.00 & 0.862 & 97 & 2 \\
21 & 13:37:09.55 & -29:52:26.87 & 320$\times$230 & 15.5 & 14.7 &  0.5 & 72.4 & 40.4 & 0.0939 & 0.297 & 5.00 & 0.303 & 326 & 7 \\
22 & 13:37:10.49 & -29:52:24.87 & 220$\times$320 & 13.0 & 9.7 &  0.6 & 3.8 & 21.4 & 0.0848 & 0.165 & 4.60 & 0.801 & 248 & 10 \\
23 & 13:37:11.07 & -29:52:30.76 & 110$\times$110 & 12.5 & 12.0 &  0.8 & 4.6 & --- & 0.0979 & 0.080 & 4.95 & 0.510 & 49 & 0 \\
24 & 13:37:09.01 & -29:52:37.80 & 190$\times$250 & 8.0 & 5.5 &  0.5 & --- & 3.4 & 0.0690 & 0.079 & 4.00 & 0.826 & 108 & 3 \\
25 & 13:37:08.42 & -29:52:50.52 & 180$\times$190 & 11.5 & 5.7 &  1.0 & 3.2 & 3.6 & 0.0497 & 0.000 & 3.60 & 0.627 & 35 & 1 \\
26 & 13:37:11.38 & -29:52:48.67 & 180$\times$260 & 7.5 & 14.6 &  1.1 & 4.6 & 23.3 & 0.0991 & 0.463 & 4.50 & 0.342 & 224 & 7 \\
27 & 13:37:10.73 & -29:52:52.56 & 190$\times$160 & 8.0 & 6.6 &  0.4 & 4.4 & 15.3 & 0.1032 & 0.157 & 4.50 & 0.577 & 117 & 6 \\
28 & 13:37:10.09 & -29:52:53.76 & 160$\times$330 & 12.0 & 10.1 &  0.6 & --- & 5.2 & 0.0612 & 0.214 & 4.25 & 1.017 & 126 & 1 \\
29 & 13:37:11.31 & -29:53:14.47 & 180$\times$200 & 27.5 & 26.0 &  1.1 & 102.3 & 52.5 & 0.0659 & 0.625 & 5.00 & 0.243 & 153 & 1 \\
30 & 13:37:10.48 & -29:53:14.32 & 240$\times$310 & 19.0 & 22.3 &  0.8 & --- & 55.1 & 0.0632 & 0.583 & 5.35 & 0.317 & 253 & 2 \\
31 & 13:37:09.57 & -29:53:16.20 & 270$\times$510 & 17.0 & 14.8 &  1.2 & 4.4 & 12.7 & 0.0632 & 0.404 & 4.50 & 0.487 & 401 & 4 \\
32 & 13:37:08.02 & -29:53:22.28 & 240$\times$380 & 22.5 & 26.7 &  3.3 & 57.5 & 44.7 & 0.0432 & 0.618 & 5.00 & 0.306 & 148 & 3 \\
33 & 13:37:11.87 & -29:53:41.79 & 370$\times$270 & 25.0 & 18.8 &  0.9 & 4.4 & 72.4 & 0.0690 & 0.812 & 5.10 & 0.217 & 375 & 2 \\
34 & 13:37:10.65 & -29:53:40.31 & 300$\times$330 & 21.5 & 14.4 &  0.3 & 4.4 & 25.2 & 0.0750 & 0.418 & 4.65 & 0.393 & 425 & 6 \\
35 & 13:37:09.54 & -29:53:34.48 & 260$\times$230 & 10.5 & 10.1 &  0.3 & 1.1 & 49.3 & 0.0576 & 0.377 & 4.10 & 0.782 & 147 & 2 \\
36 & 13:37:07.68 & -29:53:43.08 & 390$\times$490 & 25.0 & 30.8 &  4.4 & 114.8 & 51.3 & 0.0450 & 0.947 & 5.10 & 0.286 & 309 & 2 \\
37 & 13:37:08.93 & -29:53:51.84 & 310$\times$510 & 25.0 & 28.5 &  2.2 & 26.3 & 64.1 & 0.0443 & 1.865 & 5.35 & 0.242 & 287 & 4 \\
38 & 13:37:09.80 & -29:53:50.39 & 180$\times$210 & 10.5 & 11.8 &  1.0 & 2.1 & 2.0 & 0.0727 & 0.610 & 4.10 & 0.520 & 127 & 1 \\
39 & 13:37:10.29 & -29:54:04.48 & 260$\times$410 & 17.0 & 20.4 &  1.0 & 24.0 & 14.9 & 0.0825 & 0.733 & 5.00 & 0.174 & 440 & 6 \\
40 & 13:37:09.39 & -29:54:07.59 & 220$\times$180 & 10.0 & 9.9 &  0.6 & 3.3 & 4.4 & 0.0681 & 0.429 & 4.50  & 0.472 & 124 & 1 \\
41 & 13:37:08.60 & -29:54:11.04 & 200$\times$300 & 15.5 & 10.4 &  0.6 & 50.1 & 17.4 & 0.0752 & 0.453 & 4.80 & 0.449 & 183 & 1 \\
42 & 13:37:04.41 & -29:54:14.20 & 230$\times$370 & 19.0 & 20.4 &  1.5 & 36.3 & --- & 0.0690 & 0.825 & 5.00 & 0.232 & 196 & 0 \\
43 & 13:37:02.92 & -29:54:13.92 & 250$\times$270 & 25.0 & 9.8 &  0.3 & 114.8 & 4.0 & 0.0516 & 0.485 & 5.20 & 0.300 & 97 & 1 \\
44 & 13:37:02.87 & -29:54:00.80 & 440$\times$260 & 14.5 & 16.4 &  1.0 & 72.4 & 5.7 & 0.0571 & 0.469 & 4.95 & 0.245 & 224 & 5 \\
45 & 13:37:01.77 & -29:53:49.24 & 190$\times$440 & 6.0 & 6.7 &  0.5 & 5.8 & 4.6 & 0.0512 & 0.179 & 4.75  & 0.899 & 130 & 6 \\
46 & 13:37:01.04 & -29:53:42.56 & 220$\times$290 & 11.5 & 10.4 &  0.5 & --- & 4.0 & 0.0404 & 0.600 & 4.10 & 0.846 & 82 & 1 \\
47 & 13:37:02.99 & -29:53:34.17 & 250$\times$310 & 11.0 & 12.4 &  1.2 & 72.4 & 33.1 & 0.0455 & 0.333 & 5.00 & 0.427 & 120 & 2 \\
48$^{e}$ & 13:37:01.86 & -29:53:20.83 & 270$\times$290 & --- & 6.2 &  0.5 & --- & 5.0 & 0.0419 & 0.242 & 5.00 & 0.406 & 38 & 83 \\
49 & 13:37:00.79 & -29:53:19.27 & 330$\times$220 & 10.0 & 9.2 &  0.2 & 102.3 & 12.6 & 0.0333 & 0.433 & 5.25 & 0.507 & 55 & 10 \\
50 & 13:37:01.24 & -29:53:09.23 & 220$\times$220 & 6.0 & 8.0 &  0.5 & 5.8 & 27.0 & 0.0469 & 0.296 & 4.35 & 0.652 & 72 & 12 \\
\hline \hline
\label{tab:50reg}
\end{longtable}

%\end{landscape}
\end{sidewaystable}

\endgroup

\clearpage

% Table 2
% Properties of Wolf-Rayet candidates
\begingroup
\begin{table}
\begin{center}
\footnotesize
{Table 2. Wolf-Rayet Star Candidates in M83}\\
\begin{tabular}{cccccccl}
\hline \hline
RA &  DEC &  $ID_{Hadfield}$ & Reg\# & $Age_{Reg}(Myr)$ &  X & Y & Comments\\
\hline
13:37:00.41 & -29:52:54.1  &  72  &   ---  &   ---  &  3874  &  606  & faint \ha \\
13:37:01.18 & -29:52:53.9 &   73  &   --- &  --- & 3618 & 609 & faint \ha \\
13:37:01.42 & -29:51:25.8 &   74  &  4  & --- & 3512 & 2821 & strong \ha \\
13:37:03.06 & -29:50:49.5  &  78 & 7 &  4.2 & 2982 & 3698 & faint \ha  \\
13:37:04.65 & -29:52:47.5  &  82  &   ---  &   --- & 2710 &   772 & strong \ha \\
% & & & & & & & bright compact cluster just on the left \\
13:37:04.65 & -29:50:58.4 &   86  &  12  & 2.4  &  2477 & 3505 & strong \ha \\
13:37:07.35 & -29:49:39.2  &  95 &  20  &  0.8  &  1513  & 3306  & strong \ha \\
13:37:07.54 &  -29:52:53.6 &   96  &  41 & 50.1 &  1552  &  620  & strong \ha \\
13:37:07.96 & -29:52:07.9  &  99  &   ---  &    ---   &   1418 & 1762 & faint \ha \\
% & & & & & & & clear star with compact \ha \\
13:37:08.25 & -29:51:13.6 & 100  &  24  &  --- &  1311 & 3116 & faint \ha \\
13:37:08.40 & -29:52:54.9 & 102  &   ---  &   --- &  1266 & 588  & strong \ha\\
13:37:08.53 & -29:52:12.0 & 103 &  35  &  1.1 &  1224 & 1659 & strong \ha \\
% & & & & & & & bright star just on the left \\
13:37:08.70 & -29:52:28.9 & 105  &  38  &  2.1  & 1173 & 1239 & strong \ha \\
% & & & & & & & bright compact cluster just on the left \\
13:37:08.91 & -29:52:05.7 & 106 &  31 & 4.4 &  1101 & 1818 & faint \ha \\
13:37:09.02 & -29:52:45.2 & 107  &  39  & 24.0 &  1067 &   825 & no \ha \\
13:37:09.80 & -29:52:36.2 & 108  &   ---  &   ---   & 820 & 1049 & strong \ha \\
13:37:10.42 & -29:51:28.0 & 109  &  26  &   4.6  &   610 & 2756  & faint \ha \\
\hline \hline
\end{tabular}
\end{center}
\end{table}

\endgroup

\end{document}